%% file: lsndpub1.tex
\documentclass[12pt]{cernprep}
\usepackage{graphicx}
\usepackage{amssymb}
\usepackage{epsfig,color,rotating}
\usepackage{epstopdf}
\begin{document}
\newcommand{\dedx}{\mbox{${\rm d}E/{\rm d}x$}}
\newcommand{\pT}{\mbox{$p_{\rm T}$}}
\newcommand{\GeVc}{\mbox{GeV/{\it c}}}
\newcommand{\MeVc}{\mbox{MeV/{\it c}}}
\newcommand{\nue}{\mbox{$\nu_{\rm e}$}}
\newcommand{\num}{\mbox{$\nu_{\mu}$}}
\newcommand{\anue}{\mbox{$\bar{\nu}_{\rm e}$}}
\newcommand{\anum}{\mbox{$\bar{\nu}_{\mu}$}}
\newcommand{\anuetoanum}{\mbox{$\bar{\nu}_{\rm e} \rightarrow \bar{\nu}_{\mu}$}}
\newcommand{\numtonue}{\mbox{$\nu_{\mu} \rightarrow \nu_{\rm e}$}}
\newcommand{\anumtoanue}{\mbox{$\bar{\nu}_{\mu} \rightarrow \bar{\nu}_{\rm e}$}}
\newcommand{\pip}{\mbox{$\pi^+$}}
\newcommand{\pim}{\mbox{$\pi^-$}}
\newcommand{\mup}{\mbox{$\mu^+$}}
\newcommand{\mum}{\mbox{$\mu^-$}}
\newcommand{\eplus}{\mbox{e$^+$}}
\newcommand{\eminus}{\mbox{e$^-$}}
\newcommand{\Cdouze}{\mbox{$^{12}$C}}
\newcommand{\Ndouze}{\mbox{$^{12}$N}}
\newcommand{\Bdouze}{\mbox{$^{12}$B}}
\newcommand{\Rgam}{\mbox{$R_{\gamma}$}}
\newcommand{\Rbeta}{\mbox{$R_{\beta}$}}
\newcommand{\gam}{\mbox{$\gamma$}}
\newcommand{\gams}{\mbox{$\gamma$'s}}
\newcommand{\signalreaction}{\mbox{$\bar{\nu}_{\rm e}$ + p $\rightarrow$ e$^+$ + n}}
\begin{titlepage}
\docnum{CERN--PH--EP--2011--174}
\date{17 October 2011}
\begin{flushright}
\end{flushright} 
\vspace{1cm}
\title{\large{Revisiting the `LSND anomaly' I: \\ 
impact  of new data}}

\begin{abstract}
This paper, together with a subsequent paper, questions the so-called `LSND anomaly': a 3.8~$\sigma$
excess of $\bar{\nu}_{\rm e}$ interactions over standard backgrounds, observed by the LSND Collaboration in a beam dump experiment with 800~MeV protons.
That excess has been interpreted as evidence 
for the \anumtoanue\ oscillation in the $\Delta m^2$ range from 0.2~eV$^2$ to 2~eV$^2$. Such a $\Delta m^2$ range is incompatible with the widely accepted model of oscillations between three light neutrino species and would  require the existence of at least one light `sterile' neutrino. In this paper, new data on pion production by protons on nuclei are presented, and four decades old data on pion production by neutrons on nuclei are recalled, that together increase significantly the estimates of standard backgrounds in the LSND experiment, and decrease the significance of the `LSND anomaly' from 3.8~$\sigma$ to 2.9~$\sigma$. In a subsequent paper, in addition the LSND Collaboration's data analysis will be questioned, rendering a further reduction of the significance of the `LSND anomaly'.
\end{abstract}

\vfill  \normalsize
\begin{center}
The HARP--CDP group  \\  

\vspace*{2mm} 

A.~Bolshakova$^1$, 
I.~Boyko$^1$, 
G.~Chelkov$^{1a}$, 
D.~Dedovitch$^1$, 
A.~Elagin$^{1b}$, 
D.~Emelyanov$^1$,
M.~Gostkin$^1$,
A.~Guskov$^1$, 
Z.~Kroumchtein$^1$, 
Yu.~Nefedov$^1$, 
K.~Nikolaev$^1$, 
A.~Zhemchugov$^1$, 
F.~Dydak$^2$, 
J.~Wotschack$^{2*}$, 
A.~De~Min$^{3c}$,
V.~Ammosov$^{4\dagger}$, 
V.~Gapienko$^4$, 
V.~Koreshev$^4$, 
A.~Semak$^4$, 
Yu.~Sviridov$^4$, 
E.~Usenko$^{4d}$, 
V.~Zaets$^4$ 
\\
 
\vspace*{8mm} 

$^1$~{\bf Joint Institute for Nuclear Research, Dubna, Russia} \\
$^2$~{\bf CERN, Geneva, Switzerland} \\ 
$^3$~{\bf Politecnico di Milano and INFN, 
Sezione di Milano-Bicocca, Milan, Italy} \\
$^4$~{\bf Institute of High Energy Physics, Protvino, Russia} \\

\vspace*{5mm}

\submitted{(To be submitted to Phys. Rev. D)}
\end{center}

\vspace*{5mm}
\rule{0.9\textwidth}{0.2mm}

\begin{footnotesize}

$^a$~Also at the Moscow Institute of Physics and Technology, Moscow, Russia 

$^b$~Now at Texas A\&M University, College Station, USA 

$^c$~On leave of absence

$^d$~Now at Institute for Nuclear Research RAS, Moscow, Russia

$^{\dagger}$~Deceased

$^*$~Corresponding author; e-mail: joerg.wotschack@cern.ch
\end{footnotesize}

\end{titlepage}


\newpage 

\section{Introduction}

In the years 1993--1998, the LSND experiment was carried out at the Los Alamos National Laboratory.
Its scientific goal was a search for \anumtoanue\ oscillations in the `appearance' mode. The neutrino
fluxes were produced by dumping 800~MeV protons into a `beam stop'. While \num , \anum\ and
\nue\  fluxes were abundant, the \anue\ flux was vanishingly small.

The LSND Collaboration claimed an excess of \anue\ interactions 
over the expectation from standard backgrounds~\cite{LSNDPRD64}. This excess was interpreted as evidence for 
the \anumtoanue\ oscillation with $\Delta m^2$ in the range from 0.2~eV$^2$ to 2~eV$^2$ and came to be known as `LSND anomaly'.
In stark conflict with the widely accepted model of oscillations of three light neutrino flavours,
the excess would require the existence of at least one light `sterile' neutrino that does not couple to
the Z boson.

Since the `LSND anomaly' calls the Standard Model of particle physics in a non-trivial way into question,
the MiniBooNE experiment at the Fermi National Accelerator Laboratory set out to check this result.
In the neutrino mode, an oscillation \numtonue\ with parameters compatible with the LSND claim was 
not seen~\cite{MiniBooNEneutrino}. However, first results from running in antineutrino mode and
searching for \anumtoanue\ led the
MiniBooNE Collaboration to conclude that their result does not rule out 
the `LSND anomaly'~\cite{MiniBooNEantineutrino} that had
indeed been observed in the antineutrino mode.

In this situation it appears worthwhile to undertake a critical review of the original results of the 
LSND experiment that gave rise to the `LSND anomaly'.
This is the subject both of this paper and of a subsequent paper~\cite{secondpaper}.

Our review of the physics and technical papers published by the LSND Collaboration led to agreement with many of their approaches and results. However, in two areas we disagree with LSND. The first area---which 
is discussed in this paper---concerns the underestimation of the \anue\ flux from
standard sources. The second area---which will be discussed in Ref.~\cite{secondpaper}---concerns
the underestimation of systematic errors in the isolation of the signal of $\sim$120 \signalreaction\ reactions with a correlated \gam\  from neutron capture out of $\sim$2100 candidate events.

Our review of the LSND claims was carried out without insider information from members of the LSND 
Collaboration\footnote{Excepting the provision by M.~Sung of a FORTRAN program with the LSND parametrization of pion production cross-sections, and of the LSND Technical Note 134 (Neutrino Fluxes for LSND, by Myungkee Sung, Rex Tayloe and David Works, May 2000).}. We came to consider this advantageous because it led us to reproduce and assess from an entirely independent point of view the steps that were taken in the LSND analysis.

In view of the importance of the LSND result for neutrino physics, the {\em Leitmotiv\/} of our work was to answer the question: did LSND provide ``Evidence for neutrino oscillations from the observation of \anue\ appearance 
in a \anum\ beam''~\cite{LSNDPRD64}? 

Our short answer is no. Our long answer is laid down in this and a subsequent paper.

\section{Neutrino fluxes in the LSND experiment}

In order to search for the oscillation \anumtoanue\ in the  `appearance' mode, a negligible flux
of \anue\ is needed amidst a large flux of \anum . The LSND experiment was designed to realize this situation as closely as possible by dumping a large number of protons of 800~MeV kinetic energy into a totally absorbing
`beam stop' made essentially of water, copper and iron.

\begin{figure*}
\begin{center}
\includegraphics[width=0.8\textwidth]{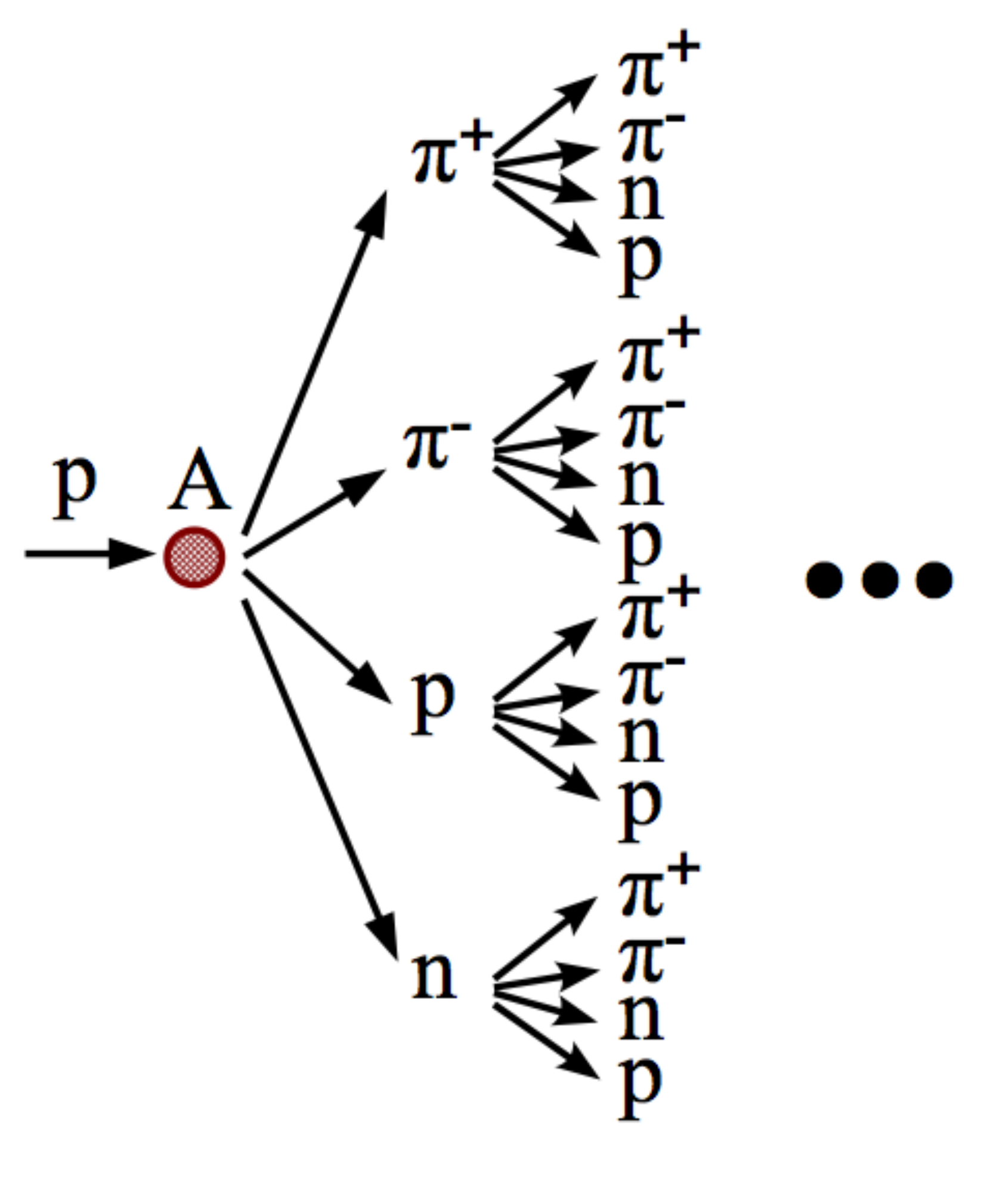}    
\caption{Schematic display of the 1st and 2nd generation hadrons in the cascade after the interaction
of an 800~MeV beam proton with a nucleus in the LSND `beam stop'.}
\label{HadronCascade}
\end{center}
\end{figure*}

The interaction of the beam protons with (primarily) O, Cu and Fe nuclei led to a hadronic cascade, schematically depicted in Fig.~\ref{HadronCascade}. 
The relatively low beam proton energy of 800~MeV assured that the hadrons in the cascade were only protons, neutrons, \pip\ and \pim , and that the number of \pip\ was an order of magnitude 
larger than the number of \pim . The totally absorbing target assured that the flight path of pions---the only short-lived particles among those hadrons---was short so that there was a small probability to decay in flight or to interact, but a large probability to come to rest after their kinetic energy was dissipated by ionization.

\begin{figure*}
\begin{center}
\includegraphics[width=0.8\textwidth]{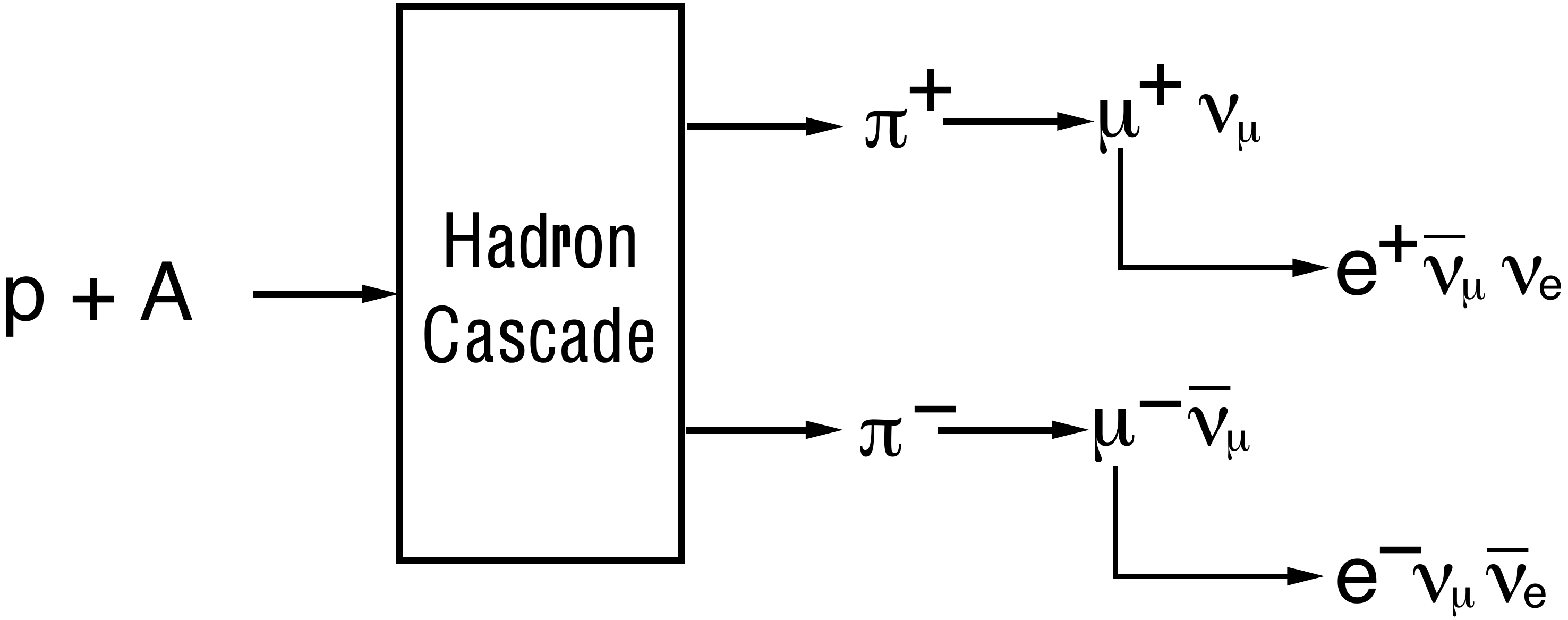}    
\caption{Schematic display of the decay processes of positive (upper chain) and negative (lower chain)
pions that are created in the hadronic cascade after the interaction of an 800~MeV beam proton with a nucleus in the LSND `beam stop'.}
\label{NeutrinosInCascade}
\end{center}
\end{figure*}

The fate of charged pions in the hadron cascade after the interaction of the beam proton are 
schematically shown in 
Fig.~\ref{NeutrinosInCascade}. Since there are decisive differences between what happens to positive and negative pions and their decay products, we discuss them in turn.

Positive pions (the upper chain in Fig.~\ref{NeutrinosInCascade}) come to rest and diffuse around until
their decay $\pi^+ \rightarrow \mu^+ \nu_{\mu}$. This creates an isotropic flux of monoenergetic \num\ of 30~MeV
and monoenergetic \mup . Because of the relatively long lifetime of muons, essentially all \mup\ will come to
rest after their kinetic energy is dissipated by ionization. They also will diffuse around until their decay
$\mu^+ \rightarrow  e^+ \nu_{\rm e} \bar{\nu}_{\mu}$. This creates isotropic fluxes of \nue\ and \anum\ with
well known respective energy spectra. Thus \pip\ lead to large fluxes of \num , \nue\ and \anum . 

By contrast, negative pions (the lower chain in Fig.~\ref{NeutrinosInCascade}) that come to rest, 
are caught in low-lying atomic orbits and disappear by strong interaction with the respective atomic nucleus, with no neutrino emitted.

Additional, albeit much less abundant, neutrino fluxes stem from pions and muons that decay in flight before they
interact or come to rest. For positive pions and muons, that means only a small addition to the fluxes of
\num , \nue\ and \anum\ from decays at rest. Of much larger importance, however, are the analogous 
fluxes of \anum , \anue\ and \num\ from the decays of negative pions and muons where the negative pion
decayed in flight.

Neutrinos from negative muon decays can either come from decays in flight or from decays at rest. In the latter
case, muon decay competes with the weak interaction of muons that are caught in low-lying atomic orbits with
the respective nucleus,  $\mu^-$ + $(A,Z)$ $\rightarrow$ $\nu_{\mu}$ + $(A, Z-1)$, that leads to an isotropic  
$\nu_{\mu}$ flux with an energy around 100~MeV which adds to the---already abundant---flux
of \num\ and therefore is of secondary interest. Of primary interest, however, is the isotropic flux of \anue\
that arises from the decay of those \mum\ in orbit that escape the capture reaction with the nucleus.
This small \anue\ flux constitutes a conventional background to the search for the \anumtoanue\ oscillation and therefore
must be well understood\footnote{In comparison, \anue\ from the decays of negative muons in flight are,
although taken into account in our analysis, nearly negligible.}.  

The conventional \anue\ flux crucially depends on
\begin{enumerate}
\item the total number of \pim\  in the hadron cascade;
\item the momentum spectra of \pim\ (the decay probability per unit length is inversely proportional to
the momentum);
\item the angular spectra of \pim\  (different drift spaces and materials are sampled at 
different angles); and
\item the geometrical layout and the material composition of the `beam stop' (in free drift spaces there is decay but virtually no
interaction and no dissipation of energy by ionization; the competition between capture and decay of
negative muons depends strongly on the atomic number $Z$).  
\end{enumerate}

As for hadron production properties, the challenge is quite daring: for protons, neutrons, and charged pions,
the interaction cross-sections and the double-differential inclusive cross-sections of secondary hadron
production are needed to calculate neutrino fluxes, in the kinetic energy range between 800 and zero MeV,
for several `beam stop' materials, primarily water, copper and iron. 

At the time of the LSND experiment, pertinent data were scarce.  

At CERN, it was realized that the HARP experiment could make a significant contribution to items 1 to 3 of the
above list, by measuring pion production in the interactions of 800~MeV (=1500~MeV/{\it c}) protons with
relevant target nuclei\footnote{We thank G.B.~Mills from the Los Alamos National Laboratory for
his suggestion to take such data.}. Being not part of the experiment's approved programme, only a few days of data taking could be set aside for the study of this issue. 

In Section~\ref{HARPexperiment} we discuss briefly those features of the HARP experiment that were used to
analyze the pertinent data. In Section~\ref{pionproduction}, we report on the cross-sections that the HARP--CDP group
extracted from these data, and that are used to cross-check the calculations of the conventional \anue\
background published by LSND~\cite{LSNDPRD64}.

\section{The HARP experiment}
\label{HARPexperiment}

The HARP experiment was designed to carry out a programme of systematic and precise (i.e., at the few per cent level) measurements of hadron production by protons and pions with momenta from 1.5 to 15~GeV/{\it c}, on a variety of target nuclei. It took data at the CERN Proton Synchrotron in 2001 and 2002. Cross-sections of large-angle pion, proton and deuteron production on a number of nuclei have been published (see Refs.~\cite{Beryllium1,Beryllium2,Tantalum,Copper,Lead,Carbon,Tin}).

For the work described here, a thin and a thick water target was exposed to a $+1.5$~GeV/{\it c} beam, with length of 6~cm and 60~cm, respectively.
The target radius was in both cases 18~mm. The water was contained in plexiglass tubes with 2~mm
wall thickness. The thin water target was used for the measurement of the double-differential inclusive
pion production cross-section by protons. The finite thickness of the thin water target leads to a small attenuation of the number of incident beam particles. The attenuation factor taken into account in the cross-section normalization is $f_{\rm att} = 0.964$.

The thick water target was used for a relative measurement: a high-statistics determination of
the $\pi^- /  \pi^+$ ratio of secondary pions. 

In addition to the water targets, data were recorded with the same thin (5\% $\lambda_{\rm int}$) tantalum, copper  and lead targets
as used for the work described in Refs.~\cite{Tantalum,Copper,Lead}.

Protons and pions with 1.5~GeV/{\it c} momentum were delivered by the T9 beam line in the East Hall of CERN's Proton Synchrotron with a momentum bite $\Delta p/p \sim 1$\%. Only the positive beam polarity was used.
The beam instrumentation, the definition of the beam particle trajectory, the cuts to select `good' beam particles, and the muon and electron contaminations of the particle beams, 
are as described, e.g., in Ref.~\cite{Beryllium1}.

Our calibration work on the HARP TPC and RPCs is described in detail in Refs.~\cite{TPCpub} and \cite{RPCpub}, and in references cited therein. The momentum resolution $\sigma (1/p_{\rm T})$ of the TPC is typically 0.2~(GeV/{\it c})$^{-1}$ and worsens towards small relative particle velocity $\beta$ and small polar angle $\theta$.
The absolute momentum scale is determined to be correct to better than 2\%, both for positively and negatively charged particles.

The polar angle $\theta$ is measured in the TPC with a 
resolution of $\sim$9~mrad, for a representative 
angle of $\theta = 60^\circ$. In addition, a multiple scattering
error must be considered that is for a proton with $p_{\rm T} = 500$~MeV/{\it c} 
in the TPC gas 
$\sim$4~mrad  at $\theta = 20^\circ$, and $\sim$12~mrad  at $\theta = 90^\circ$;
for a pion with the same characteristics, the multiple scattering errors are
$\sim$3~mrad and $\sim$6~mrad, respectively (the numbers refer to the thin water target). The polar-angle scale is correct to better than 2~mrad. 
The TPC measures \dedx\ with a resolution of 16\% for a track length of 300~mm.

The efficiency of the RPCs that surround the TPC is better than 98\%. The intrinsic time resolution of the RPCs is 127~ps and the system time-of-flight resolution (that includes the jitter of the arrival time of the beam particle at the target) is 175~ps.

\begin{figure*}
\begin{center}
\includegraphics[width=0.8\textwidth]{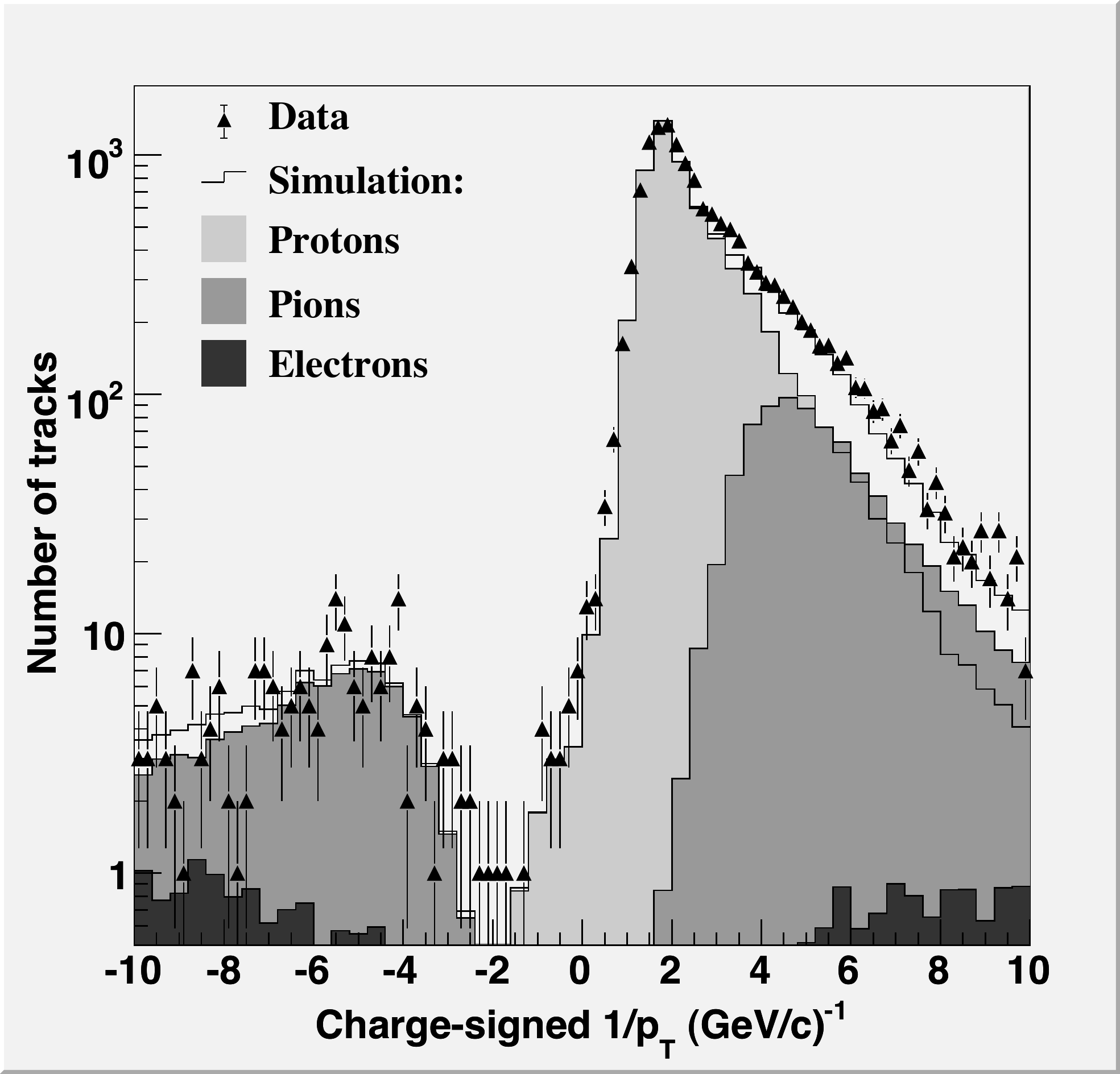}    
\caption{Statistics of  secondary protons, $\pi^+$, $\pi^-$,
e$^+$ and e$^-$ for beam protons with 800~MeV kinetic energy impinging on water.}
\label{ParticleIdentification}
\end{center}
\end{figure*}

To separate measured particles into species, we assign on the basis of \dedx\ and $\beta$ to each particle a probability of being a proton, a pion (muon), or an electron, respectively. The probabilities add up to unity, so that the number of particles is conserved. These probabilities are used for weighting when entering tracks into plots or tables.
Because of the strong preponderance of protons over $\pi^+$, great care had to be taken to separate cleanly
these two particles. To illustrate the difficulty of this undertaking, Fig.~\ref{ParticleIdentification} shows the 
statistics of secondary protons, $\pi^+$, $\pi^-$,
e$^+$ and e$^-$, with
polar angles in the range $20^\circ < \theta < 60^\circ$, for beam protons with 800~MeV kinetic energy impinging on the thin water target.

\section{Pion production by 1.5~GeV/c protons}
\label{pionproduction}

\subsection{Pion cross-sections on various target nuclei}

\begin{figure*}
\begin{center}
\begin{tabular}{cc}
\includegraphics[width=0.45\textwidth]{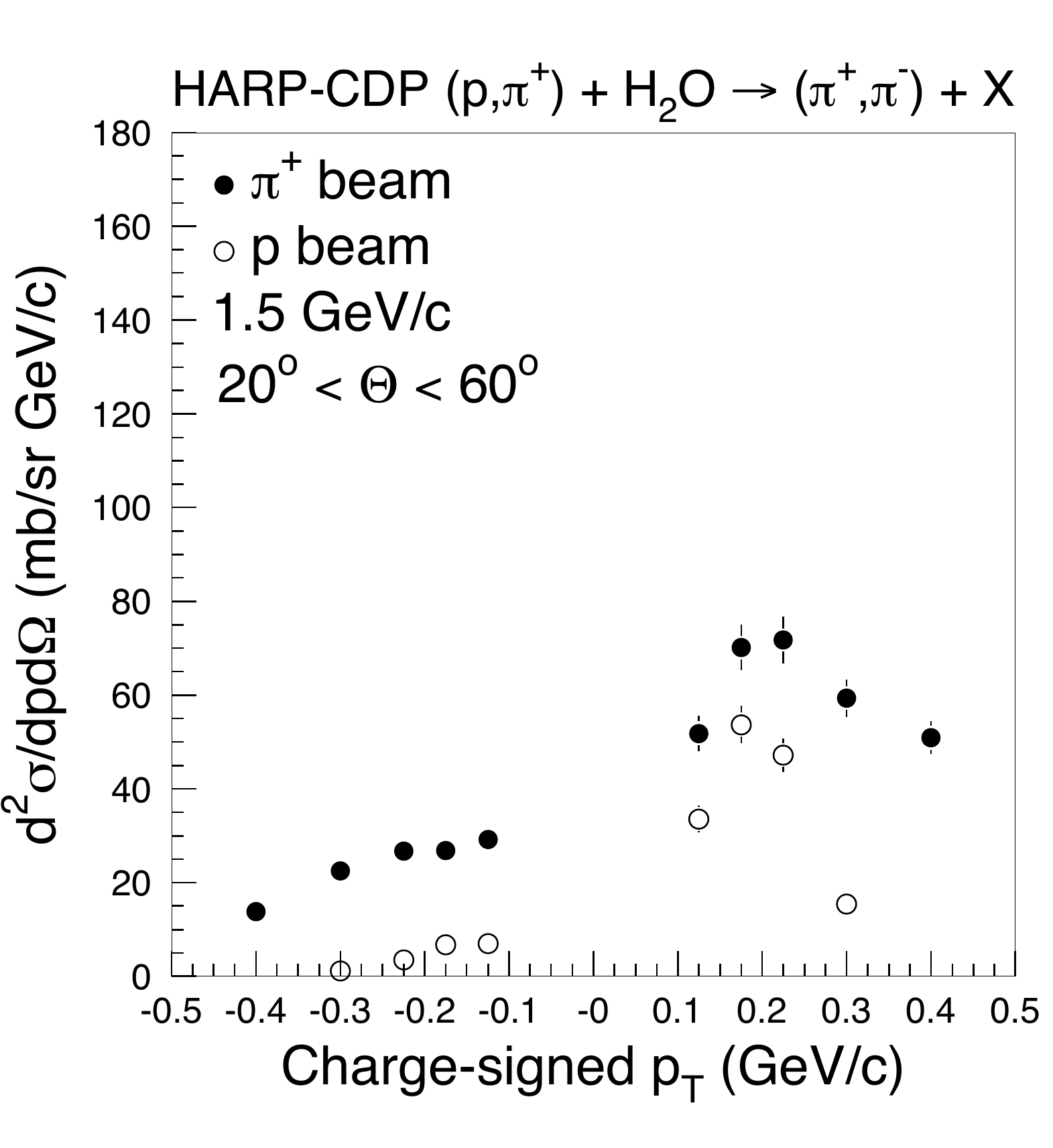}  &
\includegraphics[width=0.45\textwidth]{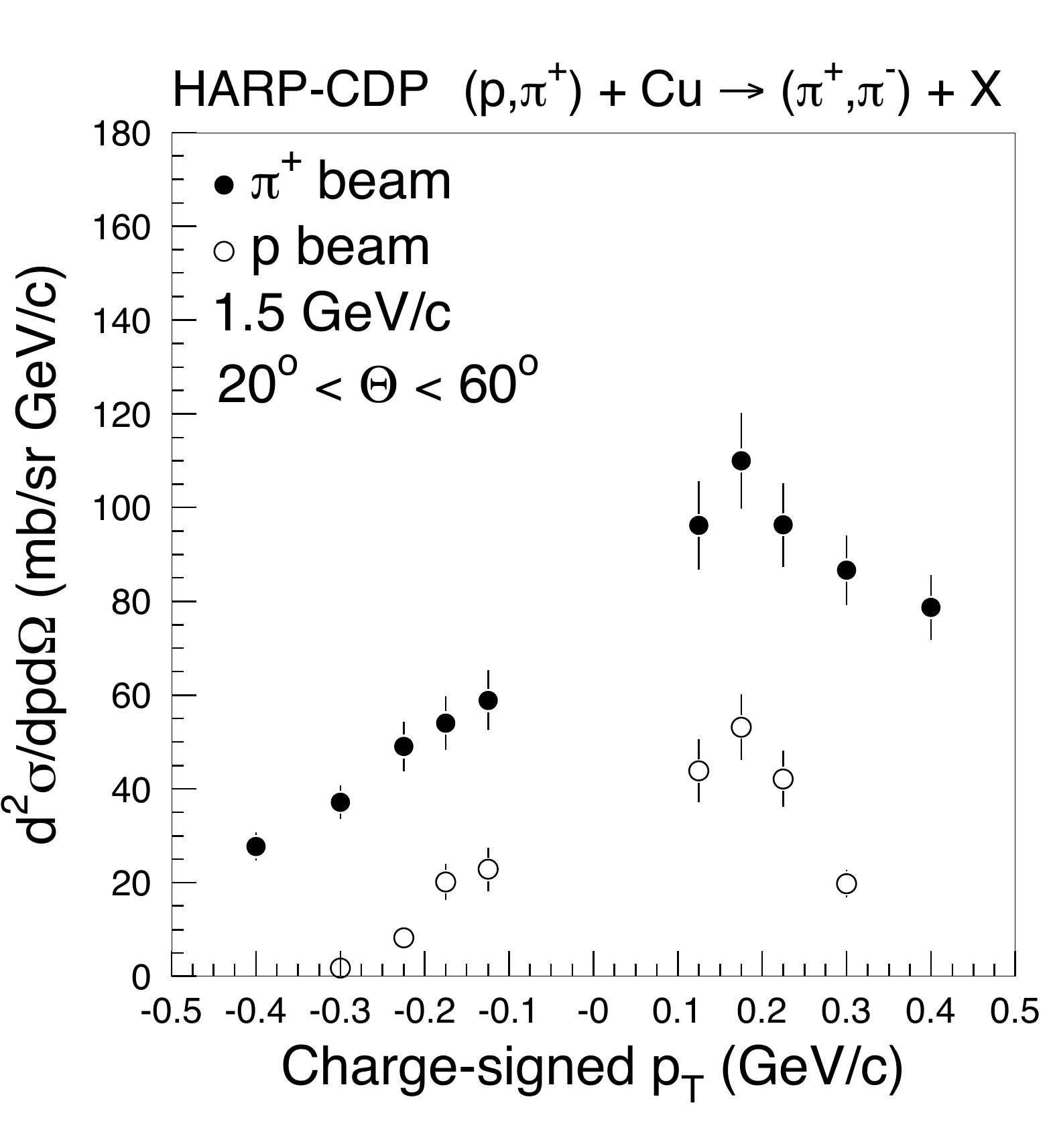}  \\
\includegraphics[width=0.45\textwidth]{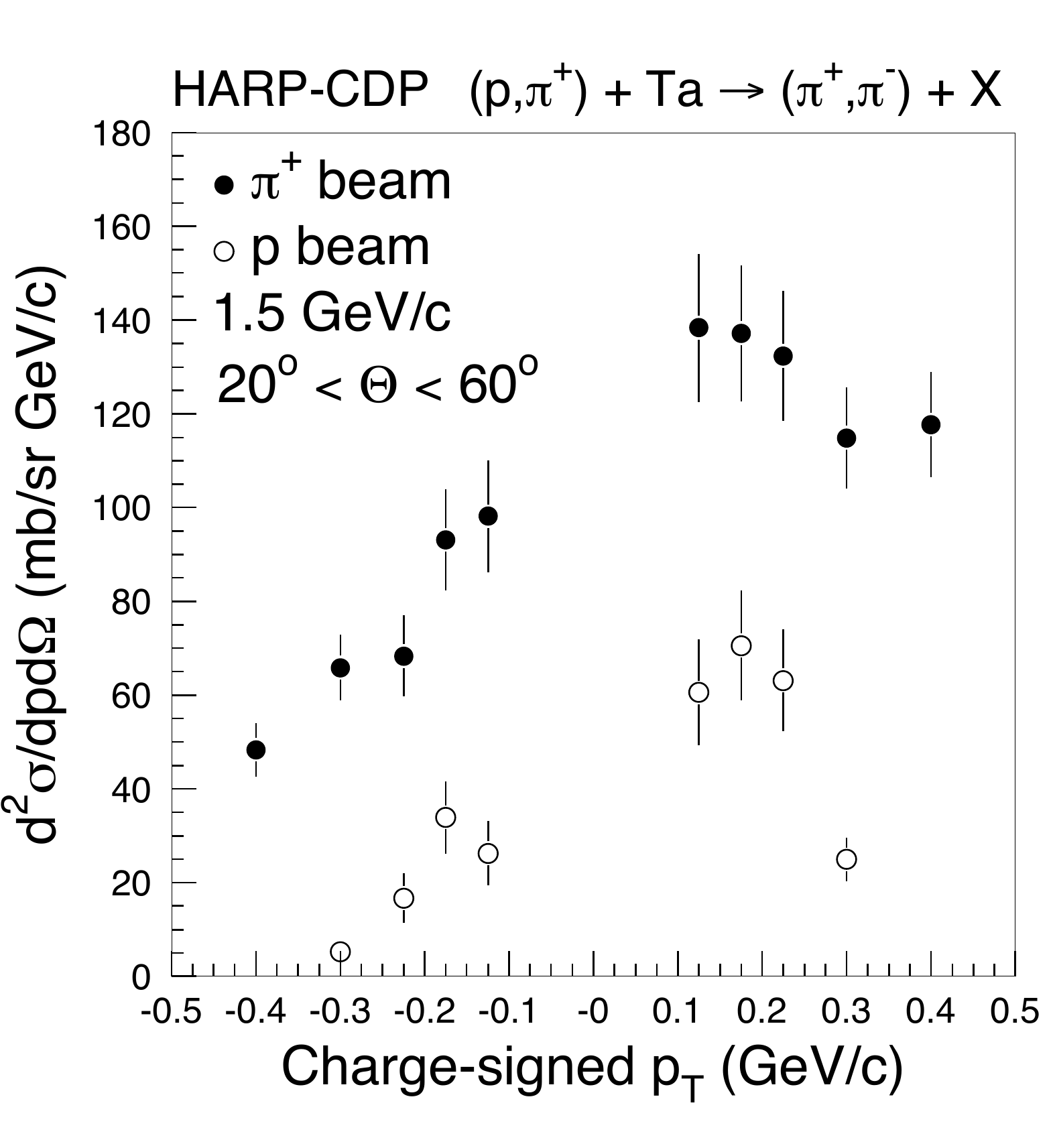}  &
\includegraphics[width=0.45\textwidth]{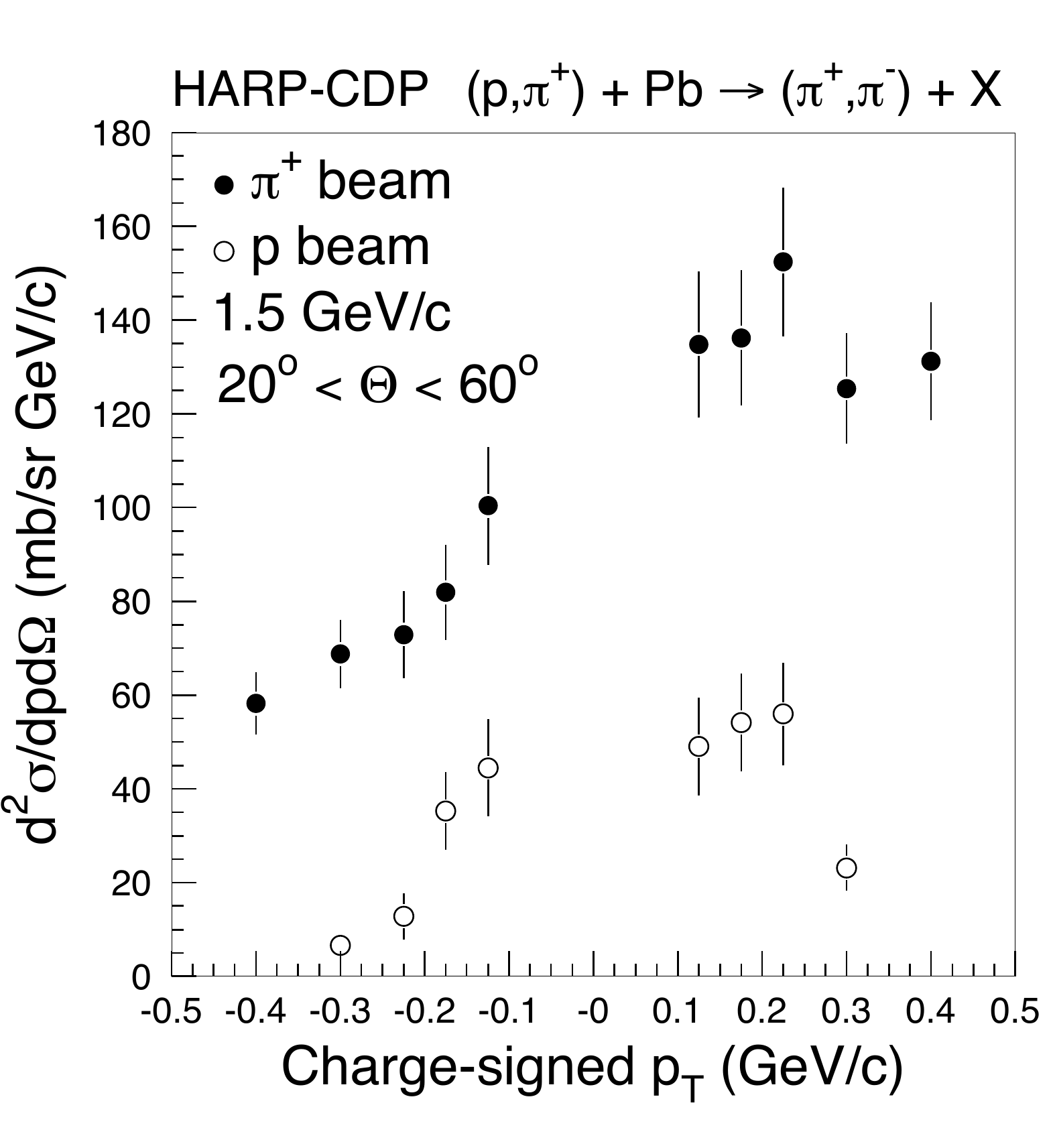}  \\
\end{tabular}
\caption{Double-differential inclusive cross-section ${\rm d}^2 \sigma /{\rm d}p{\rm d}\Omega$ 
[mb/(GeV/{\it c} sr)] of the production of $\pi^\pm$ by incoming 1.5~GeV/{\it c} 
$\pi^+$ and protons impinging on thin water (upper left panel), copper (upper right  panel),
tantalum (lower left panel) and lead (lower right panel) targets.}
\label{800MeVH2OCuTaPb}
\end{center}
\end{figure*}

In Tables~\ref{pro.H2O}--\ref{pip.Pb}, we give
the double-differential inclusive cross-sections ${\rm d}^2 \sigma / {\rm d} p {\rm d} \Omega$ of $\pi^+$
and $\pi^-$ production in thin targets of water, copper, tantalum, and lead,
for incoming beam protons and beam $\pi^+$, including statistical and systematic errors. The cross-sections
are given in bins of $p_{\rm T}$, for two ranges of polar angle: $20^\circ < \theta < 60^\circ$ and
$60^\circ < \theta < 125^\circ$. The cross-sections for the polar-angle range $20^\circ < \theta < 60^\circ$
are graphically shown in Fig.~\ref{800MeVH2OCuTaPb}.
\input{Table800MeVproH2O.tex}
\input{Table800MeVpipH2O.tex}
\input{Table800MeVproCu.tex}
\input{Table800MeVpipCu.tex}
\input{Table800MeVproTa.tex}
\input{Table800MeVpipTa.tex}
\input{Table800MeVproPb.tex}
\input{Table800MeVpipPb.tex}

Because of the limited time available for data taking, the statistical precision of the reported cross-sections
is rather poor. 
Systematic errors stem primarily from uncertainties in the momentum scale and the
absorption of secondary pions in materials between 
the vertex and the sensitive TPC volume. Smaller systematic errors stem from the momentum resolution, the
time-of-flight and \dedx\ resolutions, the subtraction of secondary protons and electrons, the subtraction
of beam muons (incoming $\pi^+$ beam only), and overall normalizations.
All systematic errors are propagated into the $p_{\rm T}$ spectra of secondary pions and then added in quadrature. They add up to a systematic uncertainty of our inclusive cross-sections of $\sim$7\% for
$20^\circ < \theta < 60^\circ$, and of $\sim$11\% for
$60^\circ < \theta < 125^\circ$, respectively. 

\begin{figure*}
\begin{center}
\begin{tabular}{cc}
\includegraphics[width=0.35\textwidth]{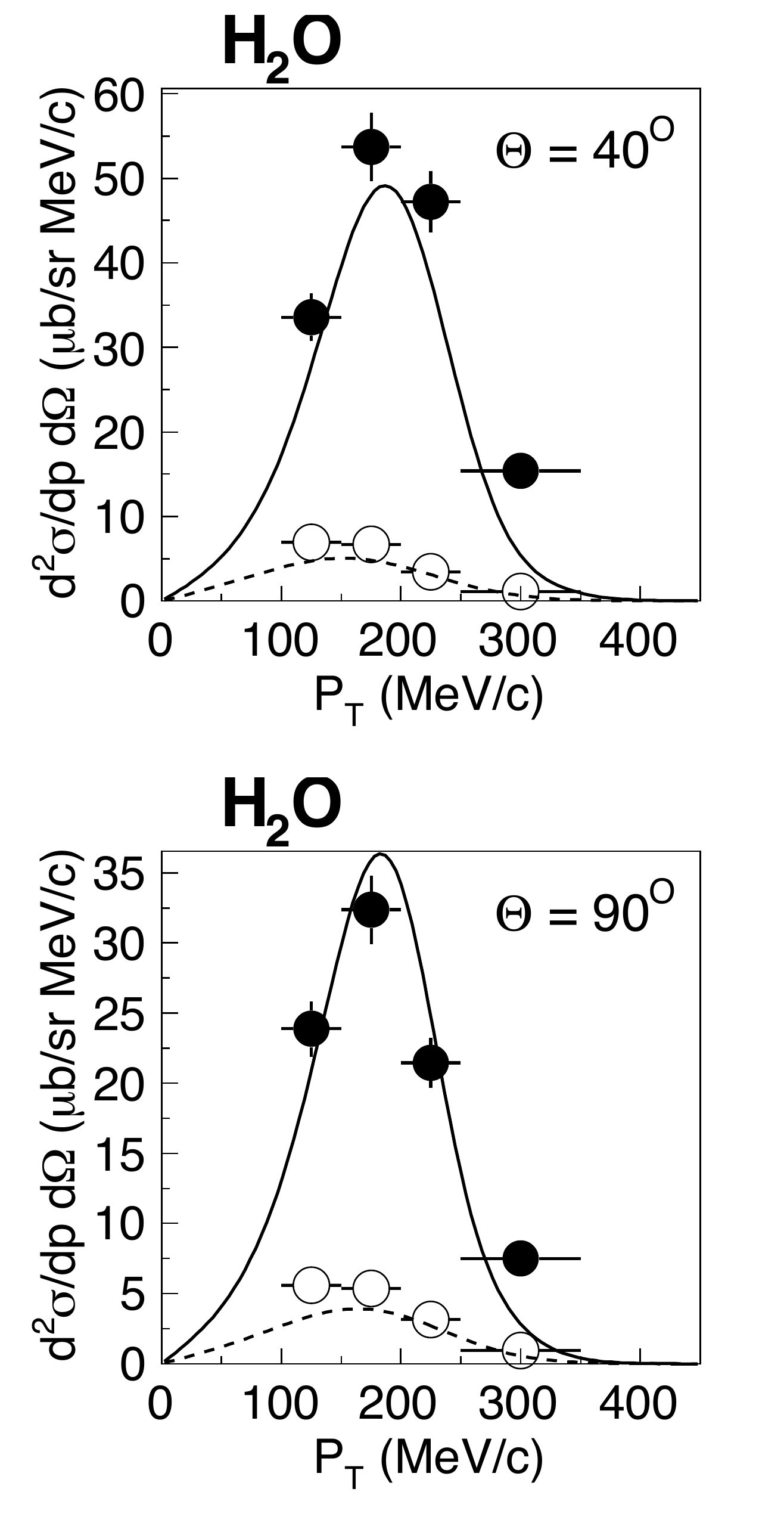} &
\includegraphics[width=0.35\textwidth]{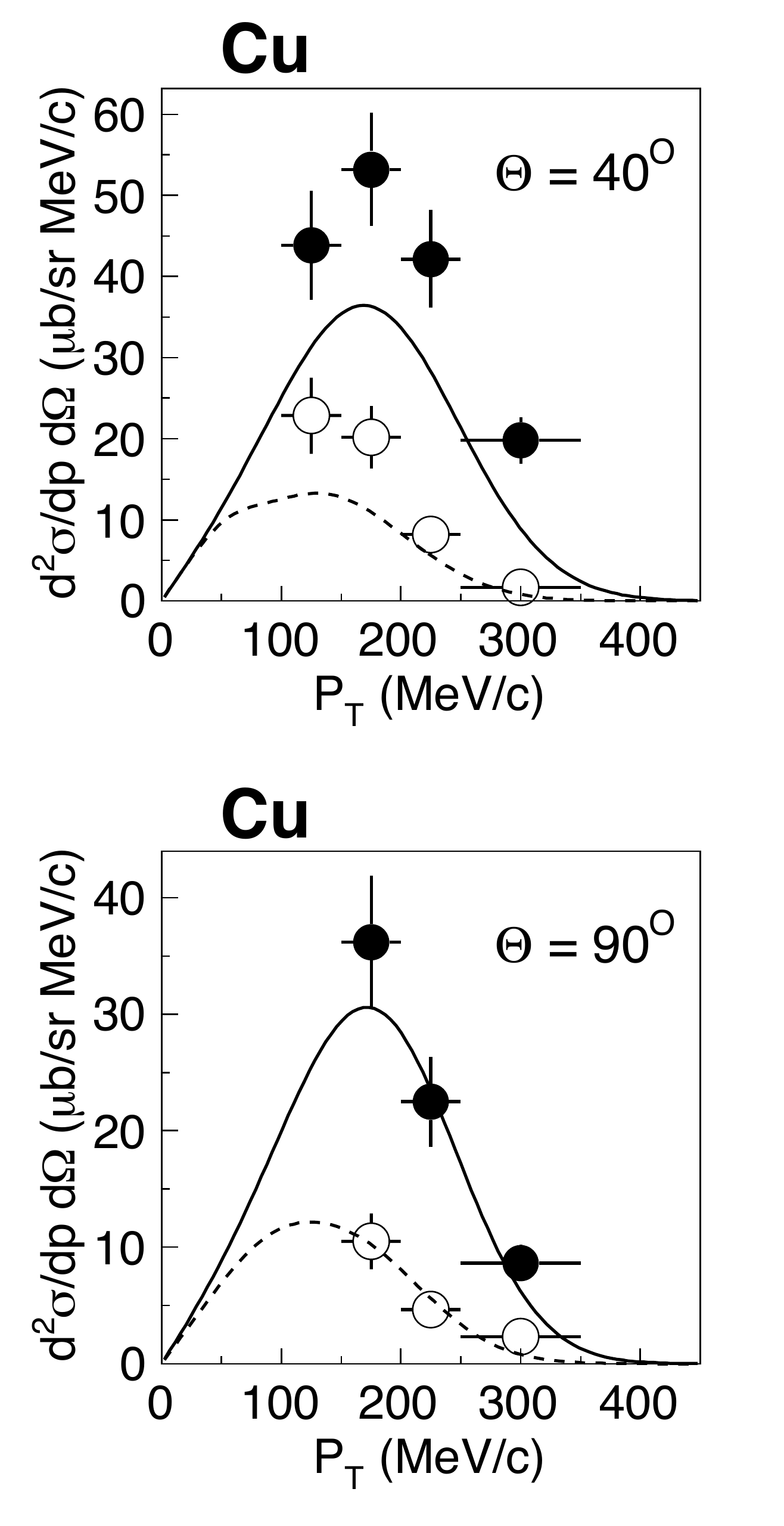} \\
\includegraphics[width=0.35\textwidth]{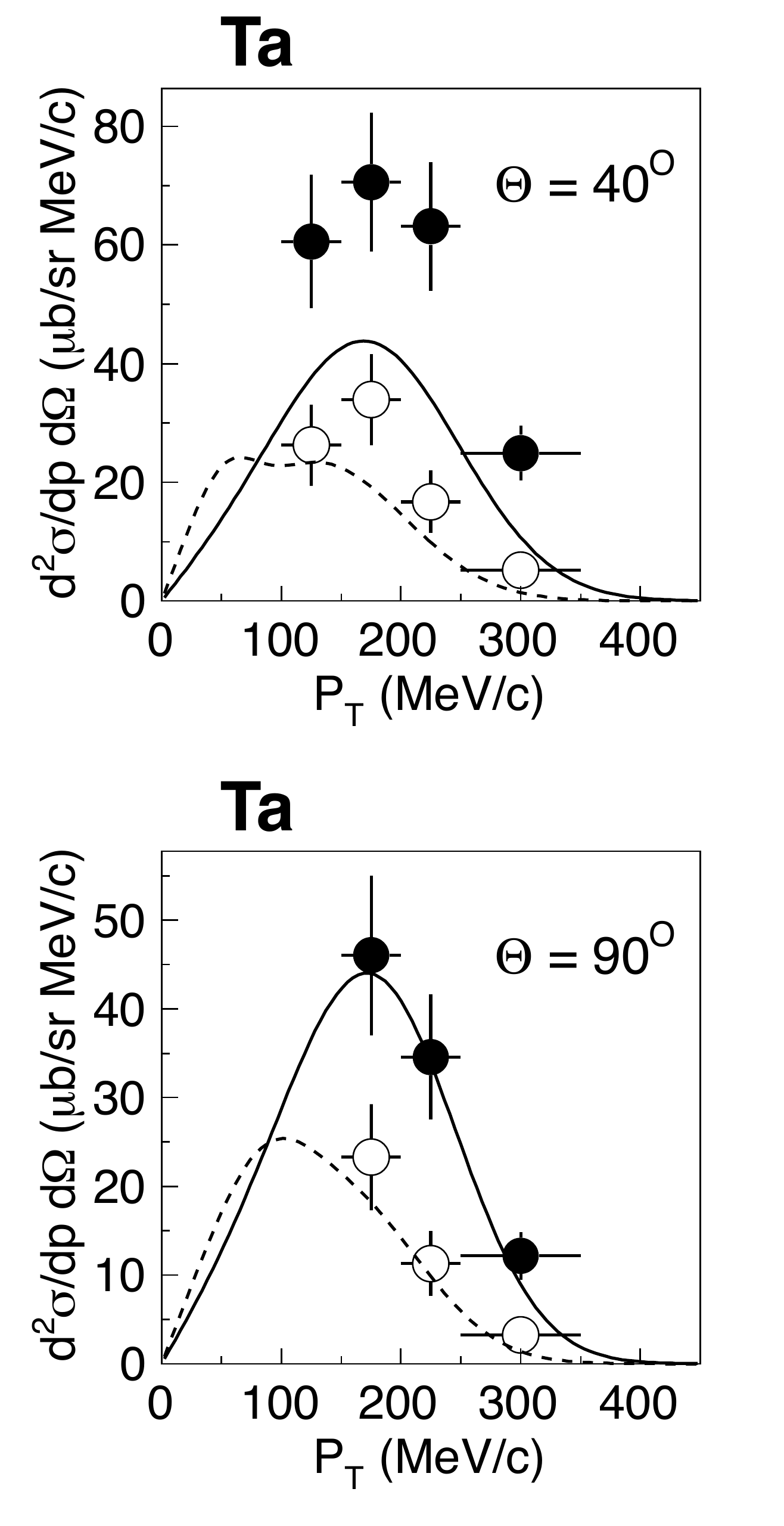} &
\includegraphics[width=0.35\textwidth]{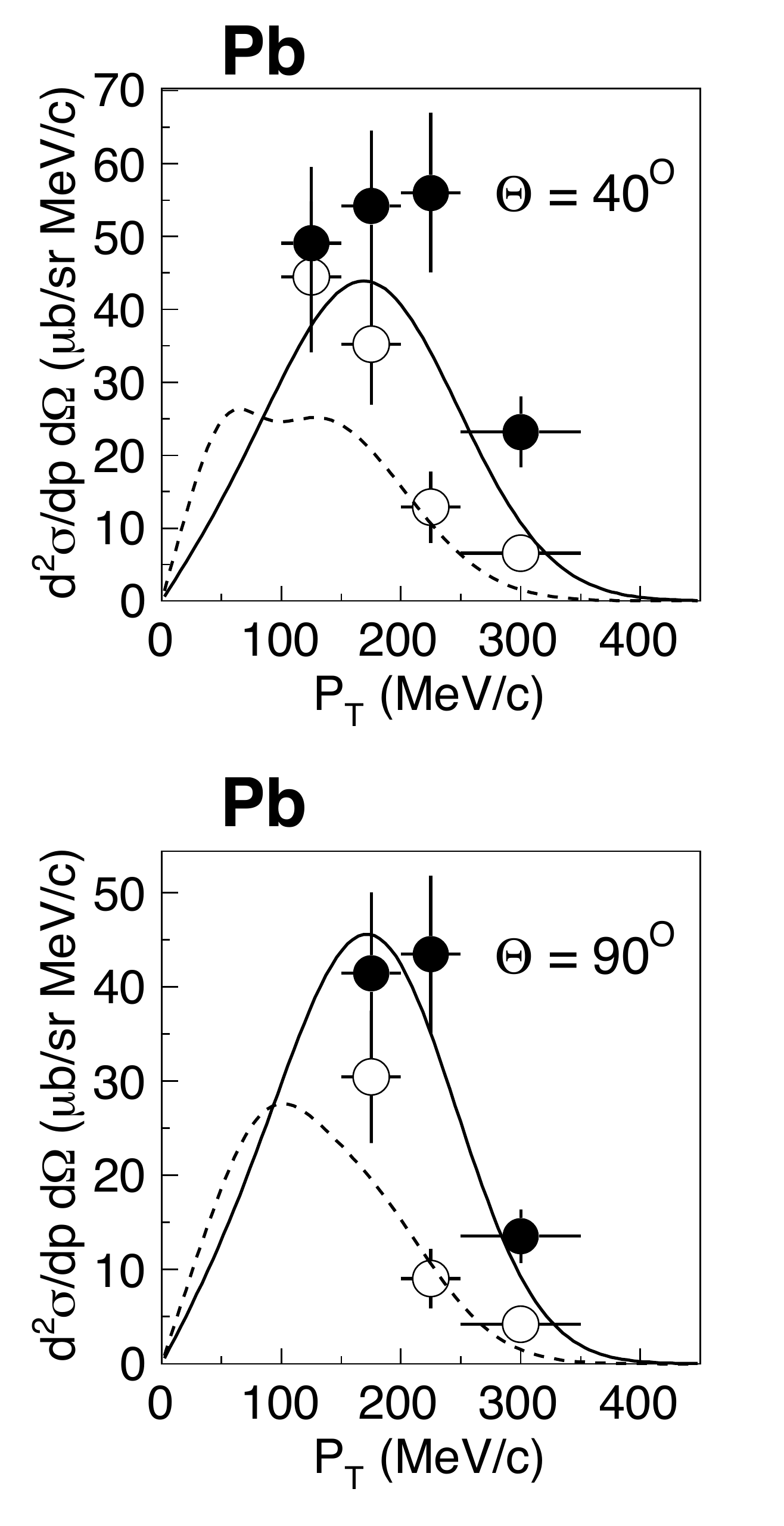} \\
\end{tabular}
\caption{Comparison of measured HARP--CDP $\pi^+$ (black points) and $\pi^-$ (open points) production 
cross-sections on water, copper,
tantalum and lead, with the ones 
predicted by the LSND parametrization (full lines for $\pi^+$, broken ones for $\pi^-$), at the polar angles of $40^\circ$  and $90^\circ$.}
\label{HARPCDPvsLSND}
\end{center}
\end{figure*}

The question that is most interesting in the context of the `LSND anomaly' is a comparison of our
measured HARP--CDP cross-sections with those that were used by LSND in the calculation of
the neutrino fluxes in their experiment. There was a considerable effort to take into account all
relevant data that existed at that time, and to merge them into what was considered by LSND a
reliable parametrization of pion production by protons on nuclei~\cite{BurmanSmith,BurmanPlischke,
BurmanSmith1989,BurmanPlischke1995}. 
Their parametrization is compared in Fig.~\ref{HARPCDPvsLSND} with our measurements. Given the
notorious uncertainty of hadronic cross-sections especially in this low-energy domain, one might argue that
the agreement is astonishingly good. In more detail, there are worrisome discrepancies visible, especially
for heavier target nuclei, where the all-important $\pi^-$ production exceeds the one 
assumed by LSND.

\subsection{$\pi^- /\pi^+$ ratio from proton interactions in water}
\label{pionratio}

Because of the importance of the $\pi^- $ production rate 
in the interactions of protons with 800~MeV kinetic energy
in the LSND `beam stop' for the prediction of the conventional \anue\ flux,  in the HARP experiment a high-statistics measurement of the $\pi^- / \pi^+$ ratio in the interactions of 1.5~GeV/{\it c} protons with
a 60~cm thick water target was undertaken. A water target was chosen since for most of the LSND data taking a thick water target contributed significantly to interactions of beam protons within the `beam stop', as will 
be detailed in Section~\ref{simulation}.

Table~\ref{ratiointhickwater} lists the ratios of inclusive double-differential
cross-sections ${\rm d}^2 \sigma /{\rm d}p{\rm d}\Omega$ of
$\pi^-$ to $\pi^+$ production by $1.5$~GeV/{\it c} protons
in the thick water target. Because of the thickness of the target, reinteractions of secondaries occur and
contribute to the observed pion yields. These ratios are deemed to be useful for the simulation of the 
hadron cascade in the LSND `beam stop' even if the measured 
$\pi^-$ to $\pi^+$ production ratio does not exactly reproduce the situation in the LSND `beam stop',
for differences in the water target geometries. 

\begin{figure*}
\begin{center}
\includegraphics[width=0.8\textwidth]{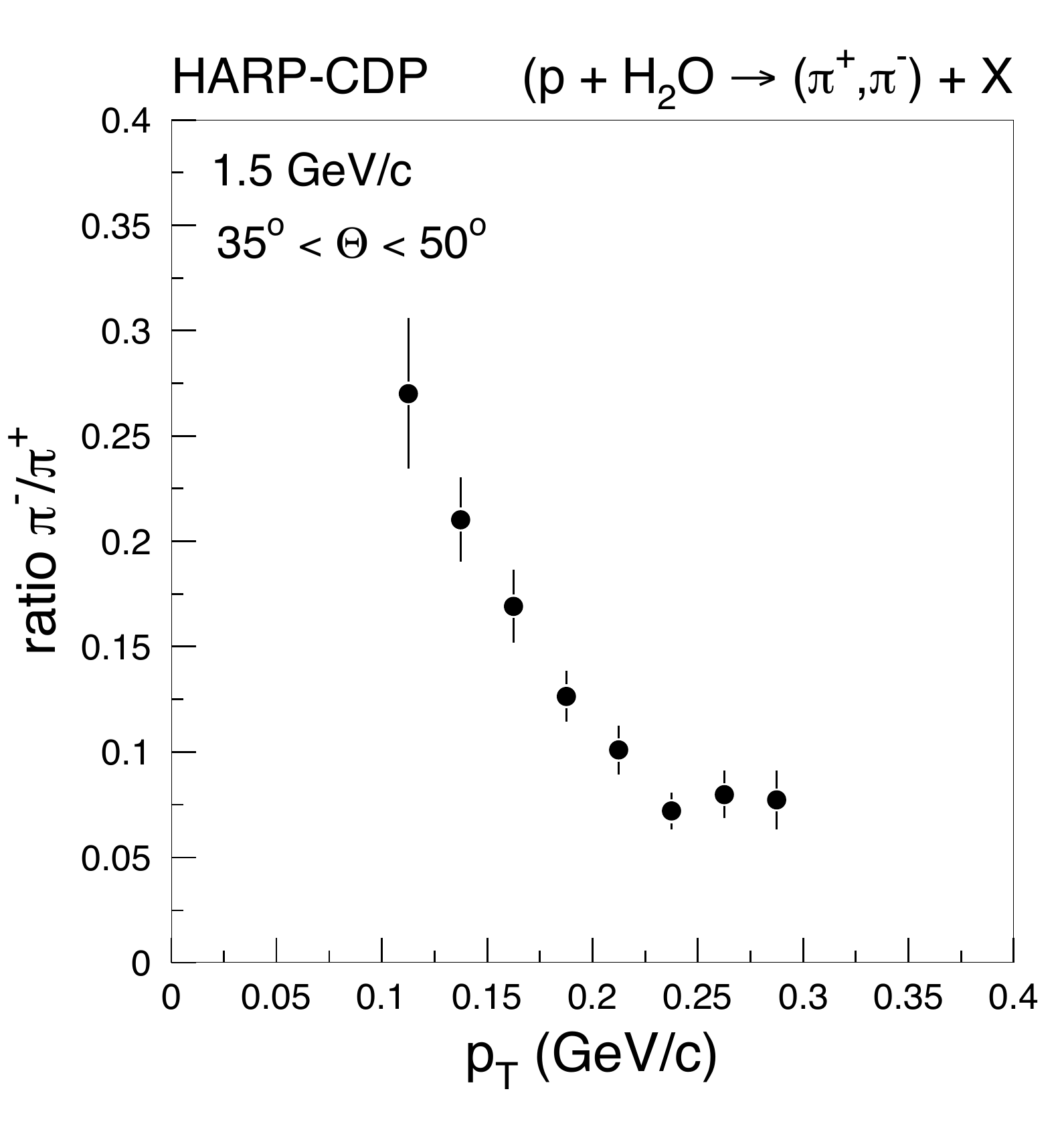}    
\caption{Ratios of inclusive double-differential
cross-sections ${\rm d}^2 \sigma /{\rm d}p{\rm d}\Omega$ of
$\pi^-$ to $\pi^+$ production by $1.5$~GeV/{\it c} protons
in the thick water target, in the polar-angle range $35^\circ < \theta < 50^\circ$.}
\label{Plot800MeVratio}
\end{center}
\end{figure*}

Figure~\ref{Plot800MeVratio} shows graphically the ratios listed in Table~\ref{ratiointhickwater}
for the polar-angle range $35^\circ < \theta < 50^\circ$.
\input{Table800MeVratio.tex}

\clearpage

\section{Pion production by neutrons}

As shown schematically in Fig.~\ref{HadronCascade}, neutrons are part of the hadron cascade 
initiated by the interaction of the beam proton with a nucleus, in much the same way 
as protons. These neutrons are vitally important for the calculation of the \pim\ content of the 
hadron cascade. While in the interactions
of $\sim$1~GeV/c protons with nuclei \pip\ secondaries strongly dominate over \pim\ secondaries,
the opposite is the case for the interactions of $\sim$1~GeV/c neutrons with nuclei. 

There are two issues relevant for the conventional \anue\ flux from \pim\ decays. The first is the 
level of neutron production within the cascade, the second is the level of \pim\ production by neutrons.
On both issues exist quite old yet relevant experimental data.

As will be discussed in detail in Section~\ref{simulation}, we base our estimations
of the neutrino fluxes originating from the LSND `beam stop' on two widely used Monte Carlo codes,
Geant4~\cite{Geant4} and FLUKA~\cite{FLUKA}. This raises the
question whether these codes reproduce correctly neutron production by protons on nuclei, 
and \pim\ production by neutrons on nuclei.

As for the first issue, neutron production by protons, we checked Geant4 and FLUKA predictions against
the results of an experiment~\cite{Baturin} that measured neutron production by 1~GeV protons on 18 nuclei
at polar angles of $4^\circ$, $7.5^\circ$ and $11.3^\circ$. While we found agreement with the
predictions of FLUKA within 15\%---what we consider satisfactory---, we found that Geant4 strongly 
underestimates neutron production by protons. 
Therefore, in certain simulation options discussed in Section~\ref{simulation}, we chose to replace Geant4 predictions of neutron production by protons by the respective FLUKA predictions.

\begin{figure*}
\begin{center}
\includegraphics[width=0.8\textwidth]{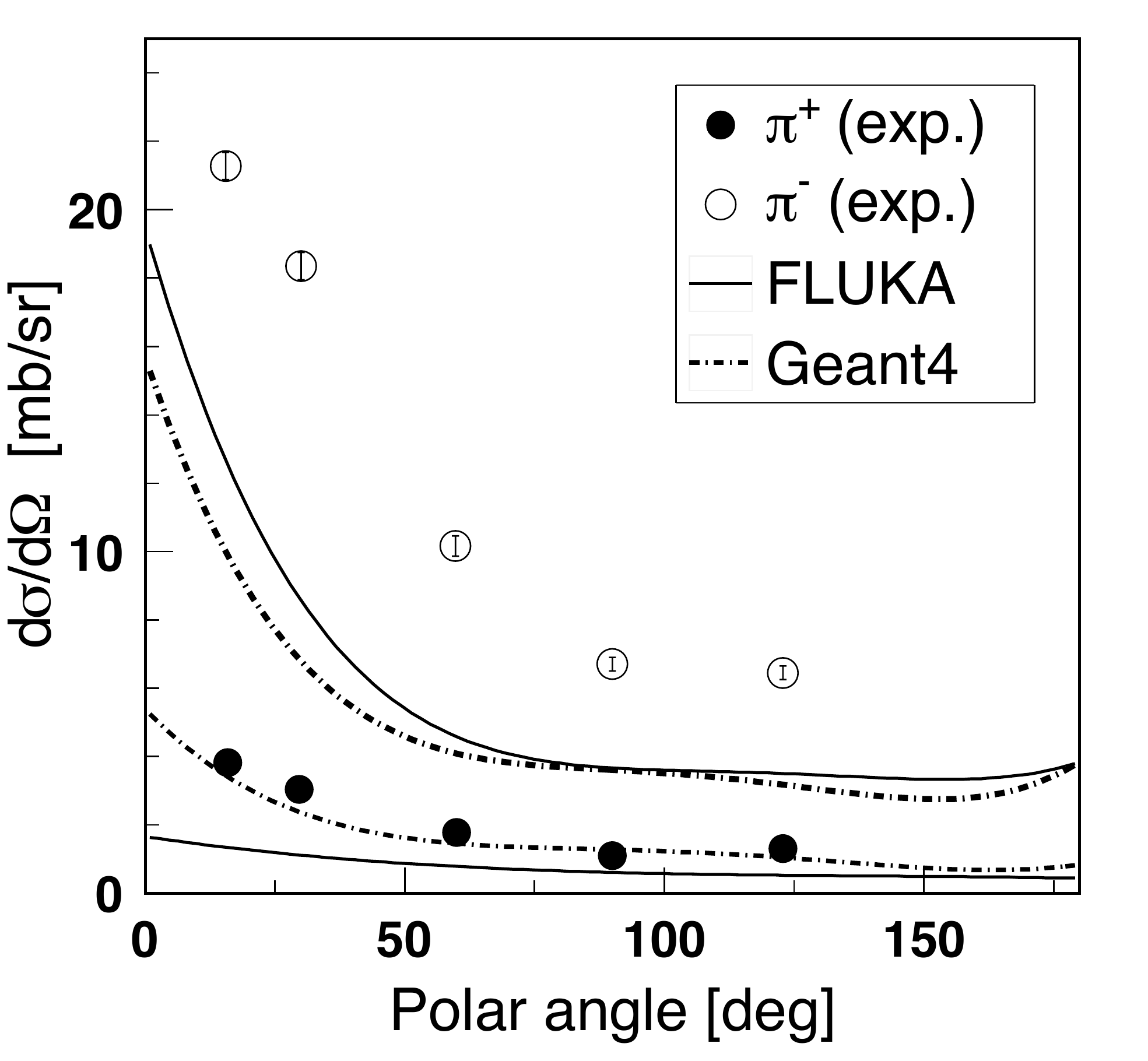}    
\caption{Comparison of inclusive cross-sections of $\pi^+$ and $\pi^-$ production by
600~MeV neutrons on copper nuclei between experiment and the predictions by FLUKA and Geant4.}
\label{Oganesianplot}
\end{center}
\end{figure*}

On the second issue, \pim\ production by neutrons, the results of a pertinent experiment~\cite{Oganesian}
are quantitatively in conflict with 
what Geant4 and FLUKA predict, although qualitatively both codes comprise pion production by 
neutrons. The situation is shown for the case of the copper nucleus in Fig.~\ref{Oganesianplot}.
We chose to ignore the small discrepancy in \pip\ production since the relative contribution of neutrons 
to \pip\ production is small, and to subsume resulting differences between FLUKA and Geant4 predictions
under our final systematic error estimate. However, the discrepancies in \pim\ production between experiment
and the Geant4 and FLUKA predictions are much too large to be ignored, given the much larger relative contribution of neutrons to \pim\ production. Therefore, in certain simulation options discussed in 
Section~\ref{simulation}, we chose to increase by a factor of two both the FLUKA and Geant4 predictions of
\pim\ production by neutrons.

To our knowledge, LSND neglected \pim\ production by neutrons which contributed to their underestimation of the
conventional \anue\ flux originating from the `beam stop'.

\section{Simulation of the neutrino fluxes in the LSND experiment}
\label{simulation}

An estimation of  the neutrino fluxes from pion and muon decays in the LSND `beam stop' is a major undertaking.
It requires a detailed simulation of the geometry and material composition of the `beam stop',
of the characteristics of the hadron cascade after the interaction of incoming beam protons, and of all
relevant physics processes in the interactions of particles with matter and in particle decays.

\begin{figure*}[h]
\begin{center}
\includegraphics[width=1.0\textwidth]{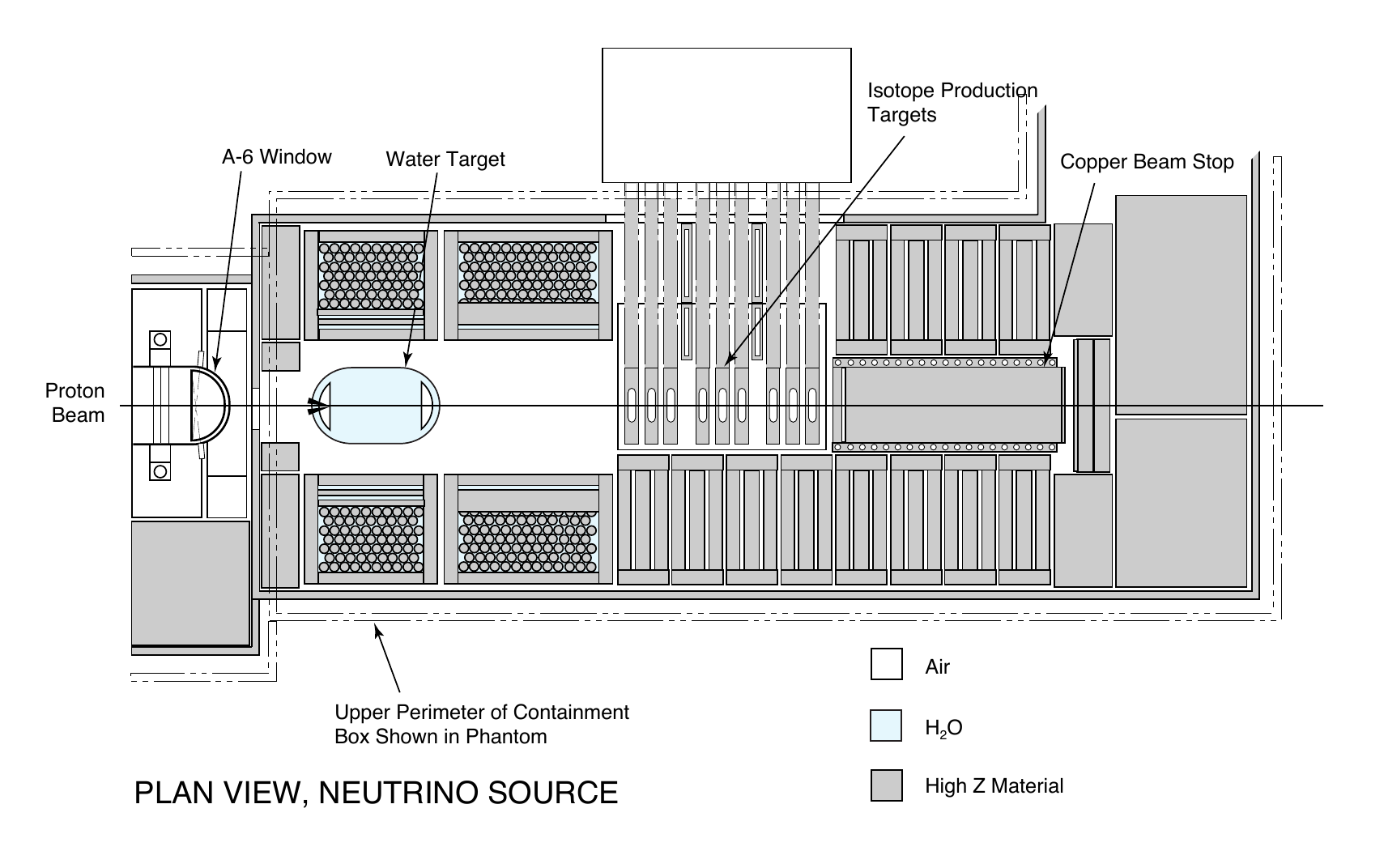}    
\caption{``The layout of the A6 beam stop, as it was configured for the 1993--1995 data taking.'' (Figure and its caption copied from Fig.~2 in Ref.~\cite{LSNDPRD64}.)}
\label{LSNDBeamStop}
\end{center}
\end{figure*}

The geometry and material composition of the LSND `beam stop' as it was configured in the years 1993--1995, is depicted in Fig.~2 of Ref. ~\cite{LSNDPRD64}
which is reproduced here in Fig.~\ref{LSNDBeamStop}. 

The three main materials of the LSND `beam stop' are the water of the primary target, 
the copper of the beam catcher, and the iron of the water target vessel and the shielding. 
Our simulation refers to the years 1993--1995 because afterwards the water target was replaced by a
``close-packed high-$Z$'' target (Section II.A in Ref.~\cite{LSNDPRD64}) which is not defined in the
necessary detail. One major uncertainty in the simulation are the ``Isotope Production Targets" that
varied considerably during the data taking. We tried to simulate on the average 
the effect of these target plates that consisted in general of rather heavy materials, by three 
molybdenum plates with 22.2~mm thickness each, spaced by 22~cm.

An intriguing feature of the LSND `beam stop' is the relatively large open space downstream of the
water target which permit more pions to decay and thus create a larger conventional \anue\
background than necessary.

With a view to keeping the likelihood of programming mistakes as low as possible, we developed two 
nearly independent Monte Carlo programs.  
One was based on the Geant4 Monte Carlo tool kit~\cite{Geant4} and focussed on a precise 
description of the geometry and material composition of the LSND `beam stop' to the best of our knowledge, 
but had limited versatility. The other was a specifically written stand-alone program, `LSNDsim', that 
employed a simplified `beam stop' geometry with reduced granularity, was fast and flexible, and hence permitted high-statistics studies of different experimental configurations. 

The only connection between our two Monte Carlo programs were the double-differential inclusive cross-sections
of secondary protons, neutrons and pions, by incoming protons, neutrons and pions. The default values of these
cross-sections were determined using the Geant4 and FLUKA programs, and implemented as options in `LSNDsim'. Within
`LSNDsim', the default values could be replaced by different ones.

The simulations of all relevant physics processes in the interactions of particles with matter and in particle decays were entirely independent.

The two programs agreed in their predictions of the conventional \anue\ flux to better than 10\%, for a closely
similar experimental configuration. We concluded that the `LSNDsim' results can be trusted. All results presented below were obtained by running `LSNDsim' with high statistics. 

`LSNDsim' provided results from the following six `core' options: 

\begin{enumerate}
\item G4 option: Geant4 cross-sections were used;
\item G4CDD option: Geant4 cross-sections were replaced by measured HARP--CDP cross-sections,
Geant4 cross-sections of neutron production by protons were replaced by FLUKA cross-sections,
and \pim\ production by neutrons was enhanced by a factor of two;
\item FL option: FLUKA cross-sections were used;
\item FLCDD option: FLUKA cross-sections were replaced by measured HARP--CDP cross-sections,
and \pim\ production by neutrons was enhanced by a factor of two;
\item LS option: cross-sections according to the LSND parametrization were used, with pions produced 
only in the primary interactions of beam protons; and
\item FLLS0 option: FLUKA cross-sections and pions from four generations were used, however with neutrons switched off and with
pions produced in the primary interactions of beam protons according to the LSND parametrization.
\end{enumerate}

Geant4 and FLUKA represent today's state-of-the-art 
in the simulation of hadronic production and hadronic interactions in matter. Differences between the
results from these two programs permits insight into systematic uncertainties.

The partial replacement of Geant4 and FLUKA cross-sections by measured values increases the reliability
of predictions of the conventional \anue\ flux. The main problem in this endeavour is that measured data 
exist only in restricted kinematical domains, for specific beam momenta and for a limited number
of target nuclei. Hence interpolations and extrapolations still could not be avoided.

HARP--CDP pion production cross-sections from the interactions of 1.5~GeV protons could be directly
used for the primary interactions of beam protons in water and copper. In those kinematic domains where
data were available, the Geant4 and FLUKA cross-sections were replaced. An average correction factor
from the measured kinematical domain was applied to correct on the average the unmeasured
kinematical domains. For the re-interactions of secondary protons of lower momentum, the same 
dependence on momentum was used throughout as implemented in the Geant4 and FLUKA cross-sections, respectively. The production of \pim\ by neutrons was enhanced by multiplying the respective Geant4 and FLUKA cross-sections by a factor of two. Pion production by \pip\ was not modified since the pertinent
HARP--CDP data are taken with incoming 1.5~GeV/{\it c} \pip\ which is too high in momentum to 
help improving the relevant cross-sections at a few 100~MeV/{\it c} momentum. 

The measured $\pi^- /\pi^+$ ratio from a 60~cm long water target reported in Section~\ref{pionratio} was used
in a fit that modified the inclusive cross-sections of \pip\ and \pim\ production by 1.5~GeV/{\it c} protons in water,
reported in Section~\ref{pionproduction}, such as to reproduce closely the measured $\pi^- /\pi^+$ ratio.

Cross-sections for the materials iron and molybdenum, not measured by HARP--CDP, were obtained
by interpolation of respective measurements for copper, tantalum and lead. For the interpolation
an $A^{2/3}$ law was used as found appropriate for pion production by low-momentum protons
(see, e.g., the discussion in Section~6 of Ref.~\cite{Carbon}).

Table~\ref{primaryprotoninteraction} lists the materials of the `beam stop' and the probability of the occurrence 
of the primary inelastic interaction of the beam proton.
\begin{table*}[htdp]
\caption{Materials of the `beam stop' and the respective probability of the primary inelastic
interaction of the beam proton therein.}
\begin{center}
\begin{tabular}{|l|c|}
\hline
Material & Probability \\
\hline
\hline
Copper (beam catcher) & 45\%  \\
Water (target)             & 27\% \\
Molybdenum (representing isotope production targets) & 22\%  \\
Iron (representing the vessel and inserts of the water target) & 5\%  \\
Aluminum (structural material) & $<1\%$ \\
Air (free space)  &  $<1\%$ \\ 
\hline
\end{tabular}
\end{center}
\label{primaryprotoninteraction}
\end{table*}

Table~\ref{piongenerations} gives a breakdown of what happens to \pip\ and \pim\ in four generations after
the primary inelastic interaction of the beam proton, for the FLCDD option of `LSNDsim'.
\begin{table*}[htdp]
\caption{Breakdown of the relative abundances of  \pip\ and \pim\ in four generations after
the primary inelastic interaction of the beam proton, and of what happens to them, 
for the FLCDD option of `LSNDsim'.}
\begin{center}
\begin{tabular}{|l|l|c|c|}
\hline
   &  & $\pip$  & $\pim$ \\
\hline
\hline
1st generation  &  & 77\% of total  & 47\% of total \\
\hline
       & Inelastic interaction & 41\%  &  34\% \\
       & Decay in flight          &  2\%   &  2\% \\
       & Decay at rest ($\pi^+$) or disappear ($\pi^-$)    &  57\%  & 64\%  \\
\hline
\hline
2nd generation &  & 20\% of total  &  40\% of total\\
\hline
       & Inelastic interaction & 32\%  &  28\% \\
       & Decay in flight          &  2\%  &  1\% \\
       & Decay at rest ($\pi^+$) or disappear ($\pi^-$)    &  66\%  & 71\%  \\
\hline
\hline
3rd generation &  & 3\% of total  & 11\% of total \\
\hline
       & Inelastic interaction & 25\%  & 26\%  \\
       & Decay in flight          & 2\%   &  1\% \\
       & Decay at rest ($\pi^+$) or disappear ($\pi^-$)    &  73\%  & 74\%  \\
\hline
\hline
4th generation  &  & $<$1\% of total & 2\% of total \\
\hline
       & Inelastic interaction & 22\%  & 24\%  \\
       & Decay in flight          &  1\%  &  1\% \\
       & Decay at rest ($\pi^+$) or disappear ($\pi^-$)    &  77\%  & 75\%  \\
\hline
\end{tabular}
\end{center}
\label{piongenerations}
\end{table*}

Negative muons that come to rest are either captured or decay in orbit. The capture probability strongly 
depends on the atomic number $Z$ but is generally well known. We took the capture 
probabilities $\lambda_{\rm capt}$ from Ref.~\cite{Suzuki}. The decay-at-rest probabilities of negative muons
$\lambda_{\rm dec}/(\lambda_{\rm dec} + \lambda_{\rm capt})$ 
that we used
in our simulation, are listed in Table~\ref{decayprobability} in decreasing order.
\begin{table*}[htdp]
\caption{Decay-at-rest probabilities  
$\lambda_{\rm dec}/(\lambda_{\rm dec} + \lambda_{\rm capt})$ of negative muons in materials of the
LSND `beam stop', as used in the `LSNDsim' program.} 
\begin{center}
\begin{tabular}{|c|c|}
\hline
Material  &  Decay-at-rest probability \\
\hline
\hline
Air &  0.862  \\
Water  & 0.818  \\
Aluminum &  0.396  \\   
Iron &  0.094  \\
Copper  & 0.074 \\
Molybdenum &  0.047 \\
\hline
\end{tabular}
\end{center}
\label{decayprobability}
\end{table*}
For the estimation of the conventional \anue\ flux, it is important that the relative weight 
of each material in which negative muons come to rest, is correctly simulated. The effect on 
different generations will be different, for their different spatial extensions and hence different 
sampling of materials. Table~\ref{anuebreakdowns} gives the fractions of the conventional \anue\ flux
that originate from \mum\ decays in different materials, and the fractions
that originate from the decays of \mum\ of different generations, for the FLCDD option of `LSNDsim'.
\begin{table*}[htdp]
\caption{Fractions of the conventional \anue\ flux
that originate from $\mum$ decays in different materials, and 
in different generations, for the FLCDD option of `LSNDsim'.}
\begin{center}
\begin{tabular}{|l|c|c|}
\hline
 \multicolumn{2}{|c|}{ } & Fraction of conventional $\bar{\nu}_{\rm e}$  \\
\hline
\hline
Material  &    \multicolumn{2}{c|}{ }   \\
\hline 
 & Iron                   &   51\%  \\
 & Copper            &   29\%  \\
 & Water               &   12 \% \\
 & Molybdenum  &   4\%  \\
 & Aluminum       &   4\%  \\
 & Air                    &   $<$1\% \\
\hline
Generation &  \multicolumn{2}{c|}{ }  \\
\hline
 & 1st                    &  74\% \\
 & 2nd                  &   22\% \\
 & 3rd                   &   4\% \\
 & 4th                    &  $<$1\% \\
\hline
\end{tabular}
\end{center}
\label{anuebreakdowns}
\end{table*}

In the following, we shall undertake the instructive comparison of the predictions for \pip\ and \pim\ production in the four options G4CDD, FLCDD, LS and FLLS0 of `LSNDsim'. We note that despite imposing HARP--CDP 
cross-section results in the G4CDD and FLCDD options, there are enough differences in the hadron production characteristics in Geant4 and FLUKA that one cannot expect the same
results from G4CDD and FLCDD. These are differences in the double-differential cross-sections of hadron production,
especially by \pim\ and neutrons where HARP--CDP have no data to contribute, and their momentum dependence;
and differences in cross-sections in kinematic regions that are outside of the regions accessible by the
HARP experiment.

\begin{figure*}
\begin{center}
\begin{tabular}{c}
\includegraphics[width=0.6\textwidth]{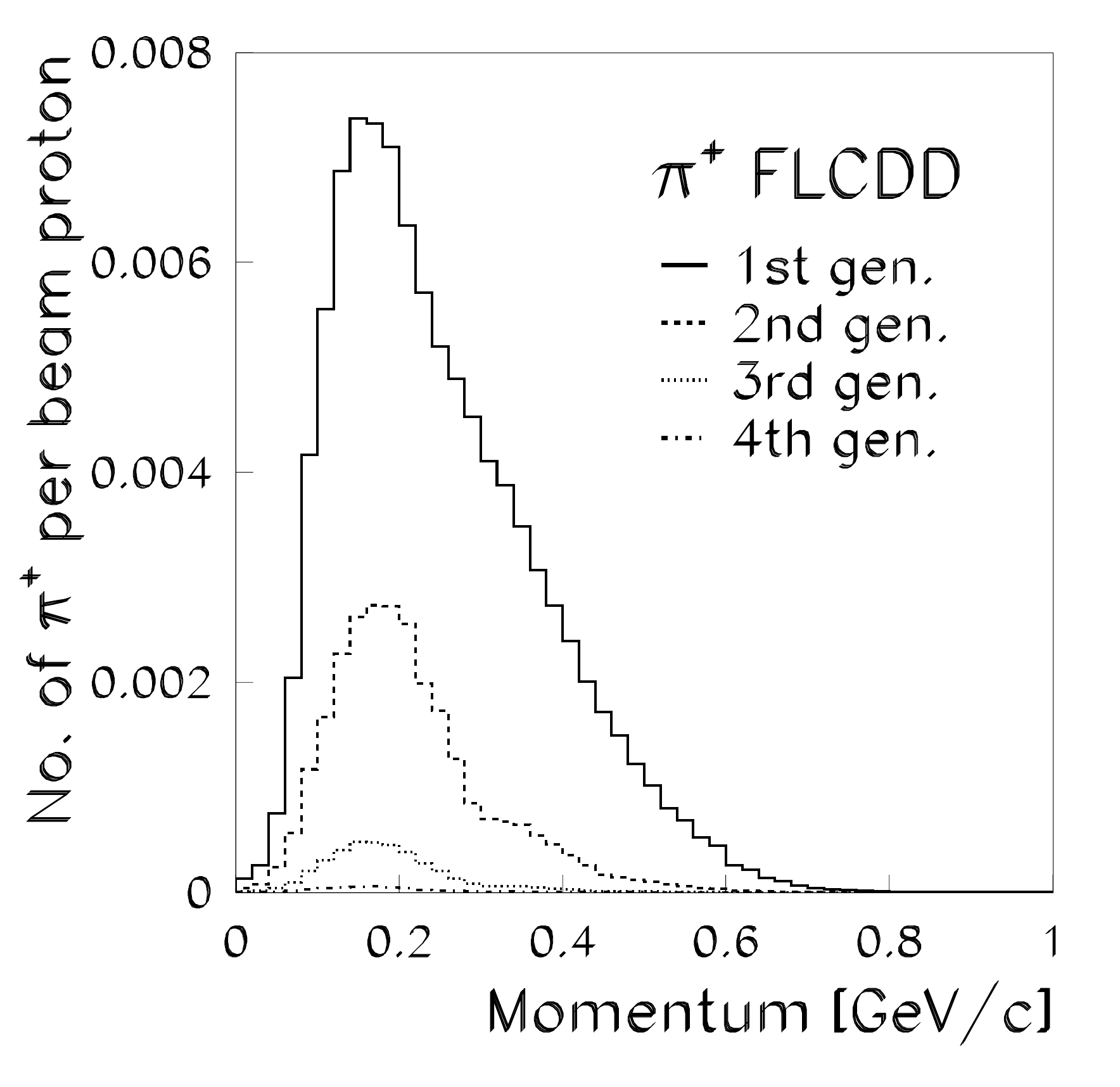} \\ 
\includegraphics[width=0.6\textwidth]{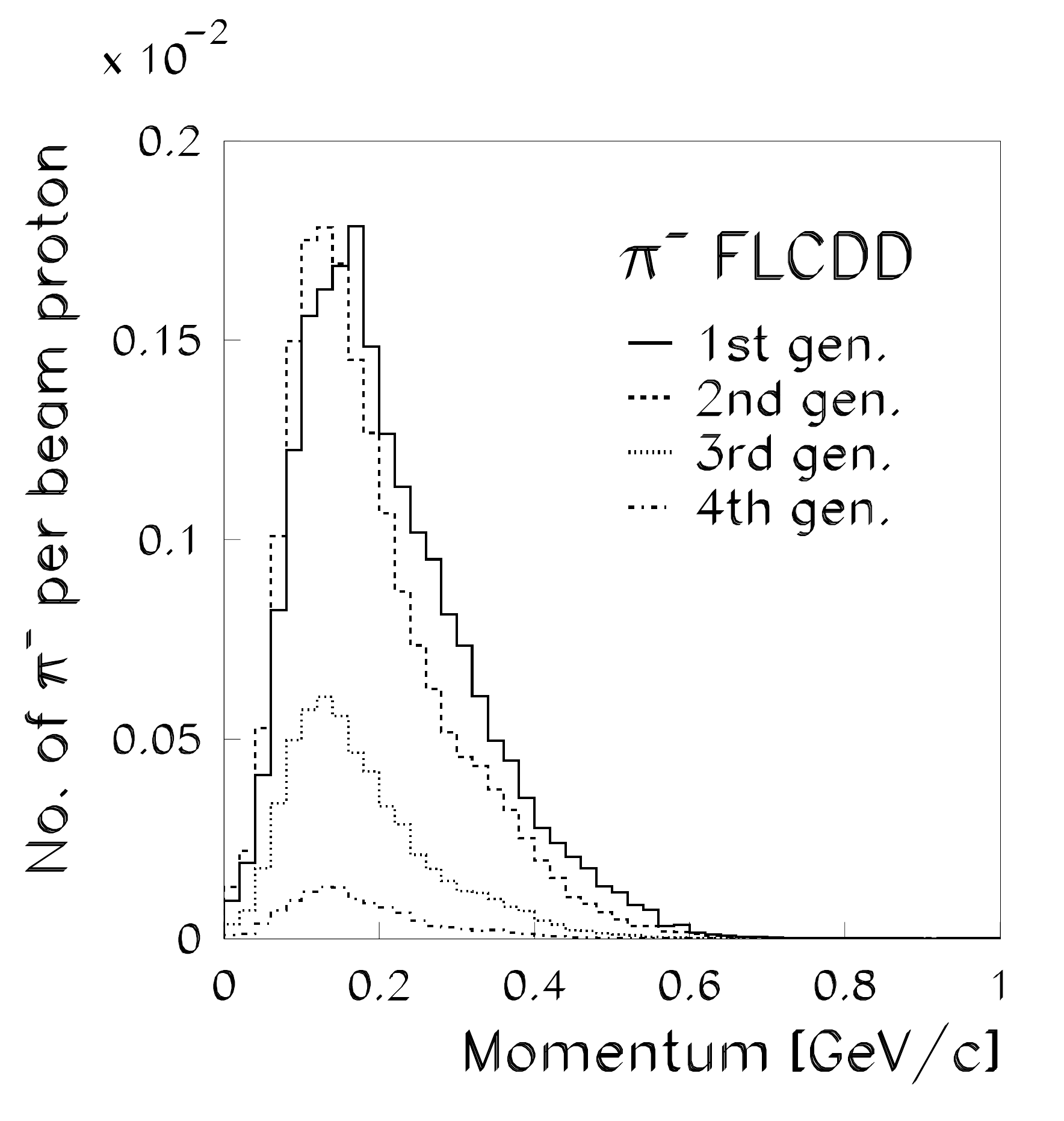} \\ 
\end{tabular}   
\caption{Absolutely normalized yields of  $\pi^+$ (upper panel) and $\pi^-$ (lower panel) production 
of the first four generations after the initial interaction of the beam proton, for the FLCDD option.}
\label{PionsinFLCDD}
\end{center}
\end{figure*}

Figure~\ref{PionsinFLCDD} demonstrates the contributions from different generations to the overall 
pion production in the LSND `beam stop',
for the example of the FLCDD option. We recall that `LSNDsim' employs four generations of hadron production 
and Fig.~\ref{PionsinFLCDD} makes clear, especially for \pim , why this was deemed necessary.  

\begin{figure*}
\begin{center}
\begin{tabular}{c}
\includegraphics[width=0.6\textwidth]{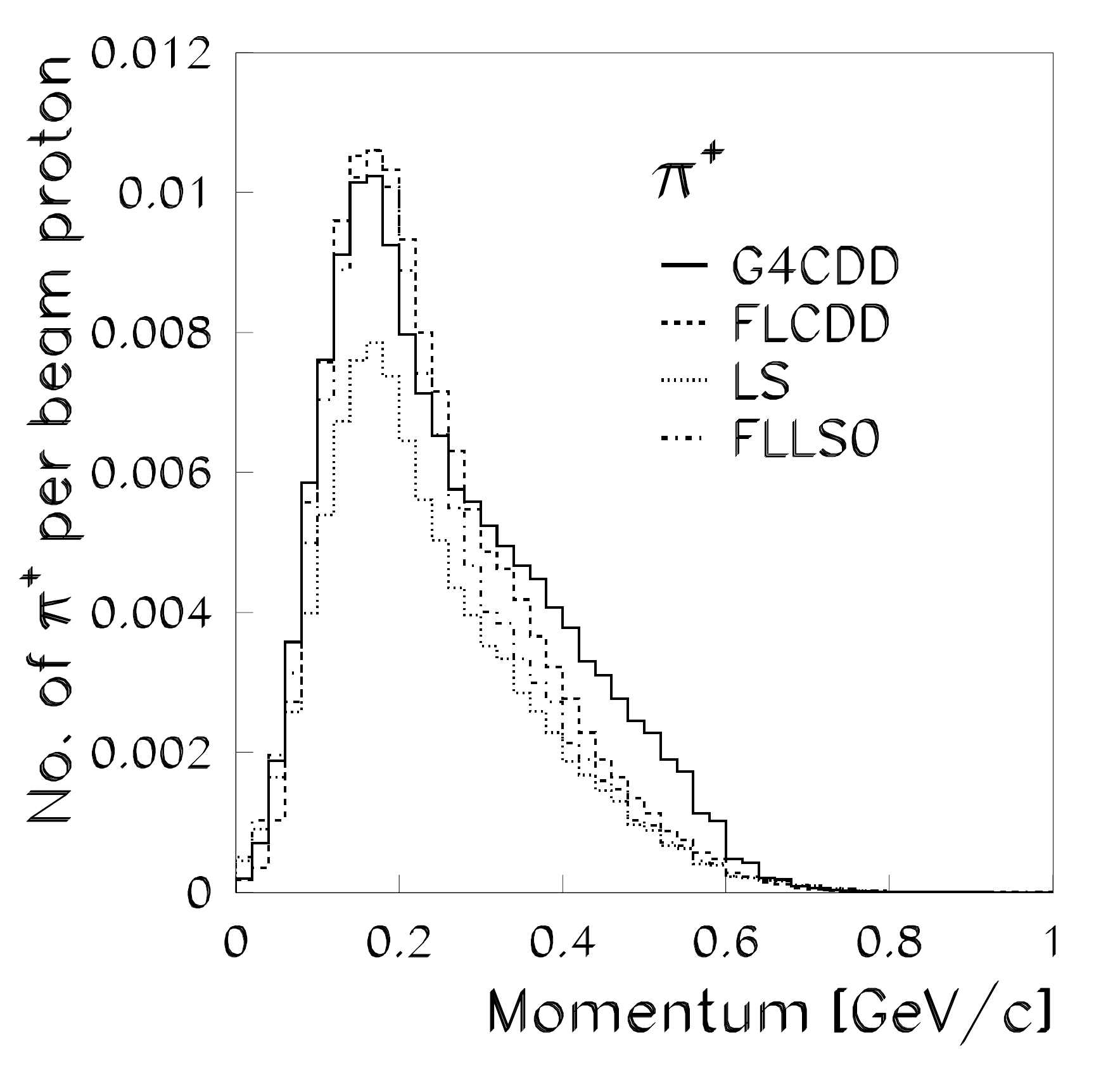} \\ 
\includegraphics[width=0.6\textwidth]{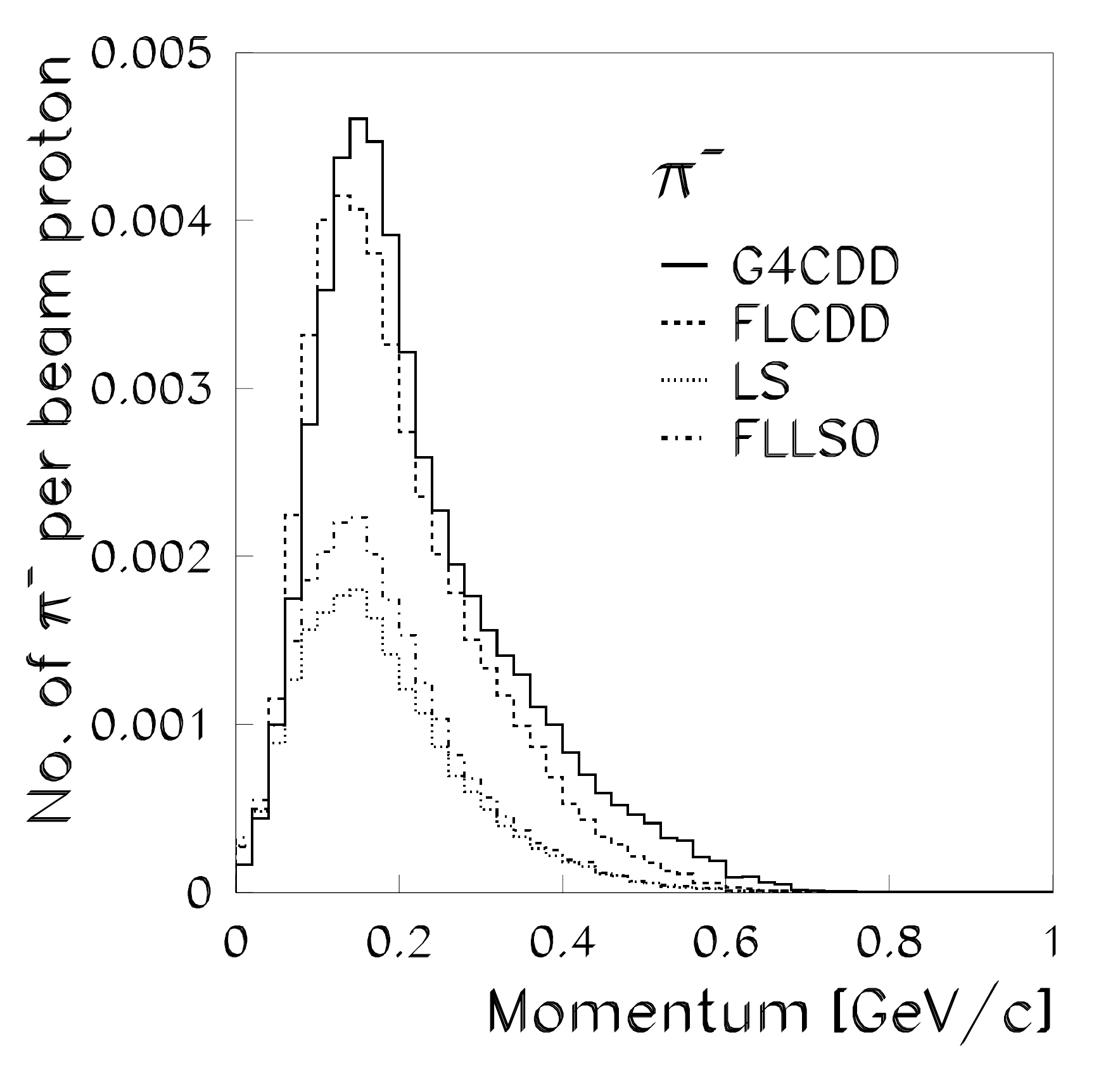} \\ 
\end{tabular}   
\caption{Absolutely normalized yields of  overall $\pi^+$ (upper panel) and $\pi^-$ (lower panel) production 
as predicted by the four options G4CDD,FLCDD, LS and FLLS0 of `LSNDsim'.}
\label{Pipandpimin4programs}
\end{center}
\end{figure*}

Figure~\ref{Pipandpimin4programs} addresses the issue of differences in the overall yields (summed over
four generations) of pion production in the LSND `beam stop' between the four options G4CDD, FLCDD, LS and FLLS0. 

While the differences in overall \pip\ production are at the 20\% level, the differences in overall \pim\ production between G4CDD and FLCDD on the one hand, and LS and FLLS0 on the other hand, amounts to a factor of two.
The cause is mainly the neglect of pion production by secondary neutrons in the LS and FLLS0 options.
At first sight this difference looks dramatic for the conventional \anue\ flux. However, it has less dramatic consequences when taking into account that pions that are produced by neutrons 
cannot occur in the first two generations, and hence such pions miss the chance of decaying in the
ample free space after the water target (c.f. Fig.~\ref{LSNDBeamStop}). Instead, their decay path is
shortened to the interaction length in the copper beam catcher, which reduces the impact of \pim\ produced
by neutrons since \anue\ arise solely from \pim\ that decay in flight before coming to rest.

Table~\ref{LSNDsummofsummstable3} presents the salient results of our simulation work and a comparison with
the results from the 1993--1995 data taking of the LSND experiment, as stated in Ref.~\cite{LSNDPRD64}.
It lists the absolute rates of \pip\ and \pim ,
the ratio $\pi^-/\pi^+$, the absolute rates of \num\ from \pip , and of \anum\ from \pim ,
that decay in flight. The \num\ and \anum\ cases hold for an energy above 123.7~MeV and 113.1~MeV
that are thresholds for the reactions $\nu_{\mu}$ + $^{12}$C $\rightarrow$ $\mu^-$ + $^{12}$N$^*$ and
$\bar{\nu}_{\mu}$ + p $\rightarrow$ $\mu^+$ + n, respectively, that lead to backgrounds as will be discussed
in Section~\ref{backgrounds}. Neutrinos of such energy can only arise from pions that decay in flight
(termed `DIF' by LSND). Further the absolute rates, from zero to the maximum energy of 
52.8~MeV,  of \anum\ and \nue , and of \num\ and \anue , are given that originate from $\mu^+$ and $\mu^-$, respectively, that decay at rest (termed
`DAR' by LSND). We recall that the rate from $\mu^+$ [DAR] decays is rather safe to calculate as it
depends essentially on the integral of all \pip\ produced in the hadronic cascade in the `beam stop'.
By contrast, the $\mu^-$ [DAR] rate (that is the centrally important one), is rather uncertain to calculate,
as it requires a \pim\ decaying in flight into $\mu^-$ + \anum , and this $\mu^-$ decaying at rest
into $e^-$ + \anue\ + \num\ in competition with much stronger nuclear capture. The ratio
$\bar{\nu}_{\rm e} / \bar{\nu}_\mu$ from muon [DAR] is given to emphasize the smallness of the conventional \anue\ flux.   
%
\begin{table*}[h]
\caption{Absolutely normalized rates of pions and neutrinos from the 1993--1995
data taking of the LSND experiment 
from several `LSNDsim' options; the rates are given per `pot' (proton-on-target) and per unit of area (cm$^2$); DIF rates refer to pion decay in flight, with the cut $E_{\nu} > 123.7 \: (113.1)$~MeV
applied; DAR rates refer to muon decay at rest, with no cut in $E_{\nu}$ applied.}
\label{LSNDsummofsummstable3}
\begin{center}
\begin{tabular}{|l||c|c|}
\hline
                         &                     &                        \\
            & LS       &  FLLS0       \\
       & (``Emulation A'')  & (``Emulation B'')   \\
     &         &                         \\
 \hline
 \hline
$\pi^+$ [pot$^{-1}]$ & 0.096        & 0.121        \\
$\pi^-$ [pot$^{-1}]$ &   0.020        &  0.024      \\
\hline
$\pi^-/\pi^+$ & 0.206 & 0.195     \\
\hline
\hline
$\pi^+$ [DIF] \num\ [$10^{-11}$ (pot$\cdot$cm$^2$)$^{-1}$]    &  2.78  &  3.10   \\
$\pi^-$ [DIF] \anum\ [$10^{-12}$ (pot$\cdot$cm$^2$)$^{-1}$]   &  2.28  & 3.08  \\
\hline
\hline
$\mu^+$ [DAR] \anum , \nue\ [$10^{-9}$ (pot$\cdot$cm$^2$)$^{-1}$]   
   & 0.520 & 0.673  \\
$\mu^-$ [DAR] \num , \anue\ [$10^{-12}$ (pot$\cdot$cm$^2$)$^{-1}$]  
   & 0.547 & 0.622 \\
\anue /\anum  &  $10.5 \times 10^{-4}$ 
   & $9.24 \times 10^{-4}$  \\
\hline
\hline 
     &                             &                         \\
     & G4             & G4CDD    \\
     &                                              &  (``Best estimate'')             \\
\hline
\hline
$\pi^+$ [pot$^{-1}]$  & 0.152  &  0.140    \\
$\pi^-$ [pot$^{-1}]$  & 0.030     &  0.050      \\
\hline
$\pi^-/\pi^+$  & 0.196 & 0.358  \\
\hline
\hline
$\pi^+$ [DIF] \num\ [$10^{-11}$ (pot$\cdot$cm$^2$)$^{-1}$]   & 6.77 &  5.57    \\
$\pi^-$ [DIF] \anum\ [$10^{-12}$ (pot$\cdot$cm$^2$)$^{-1}$]   & 5.61 & 7.83  \\
\hline
\hline
$\mu^+$ [DAR] \anum , \nue\ [$10^{-9}$ (pot$\cdot$cm$^2$)$^{-1}$]  & 0.834 
   & 0.777  \\
$\mu^-$ [DAR] \num , \anue\ [$10^{-12}$ (pot$\cdot$cm$^2$)$^{-1}$]  & 0.675 
   &  0.956  \\
\anue /\anum  & $8.09 \times 10^{-4}$ 
   & $ 12.3 \times 10^{-4}$ \\
\hline
\hline
     &                        &                       \\
     & FL         & FLCDD    \\
    &                                    &  (``Best estimate'')             \\
\hline
\hline
$\pi^+$ [pot$^{-1}]$  & 0.141   & 0.130       \\
$\pi^-$ [pot$^{-1}]$  & 0.030     & 0.045     \\
\hline
$\pi^-/\pi^+$  & 0.213 & 0.343 \\
\hline
\hline
$\pi^+$ [DIF] \num\ [$10^{-11}$ (pot$\cdot$cm$^2$)$^{-1}$]   &  2.08 & 1.83   \\
$\pi^-$ [DIF] \anum\ [$10^{-12}$ (pot$\cdot$cm$^2$)$^{-1}$]   & 2.23 &  2.44  \\
\hline
\hline
$\mu^+$ [DAR] \anum , \nue\ [$10^{-9}$ (pot$\cdot$cm$^2$)$^{-1}$]  & 0.833 
   & 0.763 \\
$\mu^-$ [DAR] \num , \anue\ [$10^{-12}$ (pot$\cdot$cm$^2$)$^{-1}$]  & 0.642
   & 0.881 \\
\anue /\anum  & $7.71 \times 10^{-4}$ 
   & $11.5 \times 10^{-4}$  \\
\hline
\end{tabular}
\end{center}
\end{table*}

As for the rationale of the comparison with LSND results we stress the following. We cannot claim that our simulation program describes the LSND `beam stop' in sufficient detail, for we rely solely on information 
that has been published. Therefore, a comparison of absolutely
normalized rates from our simulation with absolutely normalized rates published by LSND is of limited
interest. Also, in the centre of interest is not the detailed geometry of the LSND `beam  stop' and the
detailed treatment of particle transport, but rather the hadronic cascade initiated by the primary 
interaction of the beam proton. Therefore,
we replace the absolute comparison by a relative comparison where details of the `beam stop' geometry and of
particle transport cancel in first approximation. 

We claim that the `LSNDsim'
options LS and FLLS0 encompass the hadronic physics model that was employed by LSND. 
We also claim that the G4CDD and FLCDD options of `LSNDsim' represent more closely the hadronic cascade
that developed in the LSND `beam stop'.  

Our argumentation involves four steps:
\begin{enumerate}
\item we demonstrate that the LS and FLSS0 options reproduce reasonably well the results published by 
LSND (it is not required that they reproduce exactly the LSND results);
\item we take the average of LS and FLSS0 results as reference `LSND results';
\item we take the average of G4CDD and FLCDD results as the best estimate of what LSND should have
measured; 
\item we apply the difference between the average of G4CDD and FLCDD, and the average of LS and FLLS0,
to the published results of LSND and consider this as the result that LSND should have
obtained.
\end{enumerate}

In Table~\ref{comparisonLSNDwithLSNDsim}, we compare the absolutely normalized rates of pions and neutrinos from the 1993--1995
data taking of the LSND experiment as published in Ref.~\cite{LSNDPRD64}, with
the averages of the `LSNDsim' options LS and FLLS0 (`Emulation'), and G4CDD and FLCDD 
(`Best estimate').

We note the increase by a factor of 1.6 of the conventional \anue\ rate of the `best estimate', $0.919 \times 10^{-12}$~(pot $\cdot$ cm$^2$)$^{-1}$,
over  the one of the `emulation', $0.585 \times 10^{-12}$~(pot $\cdot$ cm$^2$)$^{-1}$. 
 
In our simulation, the ratio  $\pi^-/\pi^+$ refers to the average pion production in the LSND `beam stop'.
Our ratios in the LS and FLLS0 options are around 0.2, in conflict with the ratio 0.12 that is repeatedly
quoted by LSND (e.g., in Table VII of Ref.~\cite{LSNDPRD64}). We conjecture that the latter ratio must refer
to something different than the average pion production in the entire `beam stop'---that is why we put
in brackets the respective LSND number in Table~\ref{comparisonLSNDwithLSNDsim}.
\begin{table*}[h]
\caption{Absolutely normalized rates of pions and neutrinos from the 1993--1995
data taking of the LSND experiment as published in Ref.~\cite{LSNDPRD64}, and
from the average of the `LSNDsim' options LS and FLLS0 (``LSND emulation''), and G4CDD and FLCDD 
(``Best estimate''); DIF rates refer to pion decay in flight, with the cut $E_{\nu} > 123.7 \: (113.1)$~MeV
applied; DAR rates refer to muon decay at rest, with no cut in $E_{\nu}$ applied.}
\label{comparisonLSNDwithLSNDsim}
\begin{center}
\begin{tabular}{|l||c|c|c|}
\hline
    &                      &                     &                        \\
    & LSND         & LS / FLLS0       &  G4CDD / FLCDD        \\
    & published   & (``Emulation'')  & (``Best estimate'')   \\
    & (1993--1995) &         &                         \\
 \hline
 \hline
$\pi^+$ [pot$^{-1}]$ &     & 0.109    &  0.135       \\
$\pi^-$ [pot$^{-1}]$ &      &  0.022     & 0.048       \\
\hline
$\pi^-/\pi^+$ & (0.12) & 0.203 &  0.352    \\
\hline
\hline
$\pi^+$ [DIF] \num\ [$10^{-11}$ (pot$\cdot$cm$^2$)$^{-1}$]  & 1.48  &  2.94  & 3.70    \\
$\pi^-$ [DIF] \anum\ [$10^{-12}$ (pot$\cdot$cm$^2$)$^{-1}$]  & 1.57 &  2.68  &  5.14  \\
\hline
\hline
$\mu^+$ [DAR] \anum , \nue\ [$10^{-9}$ (pot$\cdot$cm$^2$)$^{-1}$] & 0.8  
   & 0.597 & 0.770  \\
$\mu^-$ [DAR] \num , \anue\ [$10^{-12}$ (pot$\cdot$cm$^2$)$^{-1}$] & 0.65  
   & 0.585 & 0.919 \\
\anue /\anum & $8.1 \times 10^{-4}$ &  $9.80 \times 10^{-4}$ 
   & $11.9 \times 10^{-4}$  \\
\hline
\end{tabular}
\end{center}
\end{table*}

Table~\ref{LSNDsimerrorestimate} lists the uncertainty estimates that we felt appropriate to assign to our
`LSNDsim' simulation.
\begin{table*}[htdp]
\caption{Estimated uncertainties in the `LSNDsim' simulation.}
\begin{center}
\begin{tabular}{|l|c|}
\hline
\multicolumn{2}{|c|}{For both the $\pi^+$ and $\pi^-$ chains}  \\
\hline   
\hline   
HARP--CDP systematics                    & 8\% \\  
Rescaling of unmeasured regions  &  3\%  \\  
`Beam stop' geometry                          & 3\%  \\
Cross-section binning                          & 6\% \\  
Cross-section interpolation                 & 6\%  \\
Hadron multiplicity                                & 5\% \\  
Extrapolations to different nuclei       &  4\% \\  
No 5th hadron generation                  &  1\% \\
\hline
\multicolumn{2}{|c|}{In addition for the $\pi^-$ chain}  \\
\hline
\hline
`LSNDsim' statistics                             & 2\% \\
$\pi^-$ production by neutrons         & 3\% \\
neutron production by protons          &3\% \\
muon capture                                       & 4\% \\
Decay space geometry                      & 5\% \\
Momentum spectrum                          & 15\% \\ 
\hline
\end{tabular}
\end{center}
\label{LSNDsimerrorestimate}
\end{table*}

To the systematic errors listed in Table~\ref{LSNDsimerrorestimate}, a 5\% error for the \pip\ chain
and a 9\% error for the \pim\ chain from HARP--CDP data statistics is to be added.

Adding all errors in quadrature, the overall error for neutrinos from the \pip\ chain is 15\%, and 24\%
for the \pim\ chain. These are the errors that we consider appropriate for our prediction of absolute integral
neutrino fluxes from the LSND `beam stop'. 

Not all listed errors are relevant when one refrains from absolute predictions and asks what
error to be assigned to the relative comparison between `LSNDsim' options LS and FLLS0 on the one hand,
and the LSND Monte Carlo on the other hand. We consider that the errors are for the $\pi^+$ chain 
not less than 9\%, and for the $\pi^-$ chain not less than 11\%, respectively.

\clearpage

\section{New estimates of Backgrounds}
\label{backgrounds}

The `LSND anomaly' is the synonym of a rate of \signalreaction\ events that is much larger than expected. 
In such a situation, a thorough discussion of backgrounds is mandatory. 

The reaction \signalreaction\ has two types of backgrounds. Background I is caused by the \anue\ flux of conventional origin. 
According to the discussion on the conventional \anue\ rate in Section~\ref{simulation}, this background is to be increased by a factor of 1.6 and its relative error increases from
20\% to 29\%.
 
Background II consists primarily of events with low-momentum muons in the final state that
are `invisible' (we remind of the copious fluxes of \num\ and \anum\ in the LSND experiment that make this
background important; normally, a muon is identified as such in the LSND detector and the respective event does not enter the sample of signal candidates). LSND state several reasons for a muon to be `invisible'. The by far most important one is a kinetic muon energy
$T_{\mu} < 3$~MeV (Section VII.A in Ref.~\cite{LSNDPRD64}).
Of prime interest is the reaction $\bar{\nu}_{\mu}$ + p $\rightarrow$ $\mu^+$ + n since it leads
to a neutron in the final state. Less important are the 
reactions $\bar{\nu}_{\mu}$ + $^{12}$C $\rightarrow$ $\mu^+$ + $^{12}$B$^*$ and
$\nu_{\mu}$ + $^{12}$C $\rightarrow$ $\mu^-$ + $^{12}$N$^*$. The common feature is that
the final-state muon is `invisible', that final-state particles (or decay
products from final-state nuclei) lead to or fake the neutron capture process, and that the event is triggered by an electron from muon decay.

The difference between LSND's and our assessment of Background II event numbers
is less in the expectation values but rather in their uncertainties.

Only \num\ with energy above 123.7~MeV can initiate the reaction 
$\nu_{\mu}$ + $^{12}$C $\rightarrow$ $\mu^-$ + $^{12}$N$^*$. The thresholds for the reactions
$\bar{\nu}_{\mu}$ + p $\rightarrow$ $\mu^+$ + n and 
$\bar{\nu}_{\mu}$ + $^{12}$C $\rightarrow$ $\mu^+$ + $^{12}$B$^*$ are 113.1~MeV and 119.7~MeV,
respectively~\cite{Audi}. That means that only the high-momentum end of the \pip\ and \pim\ spectra
can give rise to such neutrinos. These small portions
of the \pip\ and \pim\ spectra have a larger uncertainty than the integral of these spectra, though.

\begin{figure*}
\begin{center}
\includegraphics[width=0.8\textwidth]{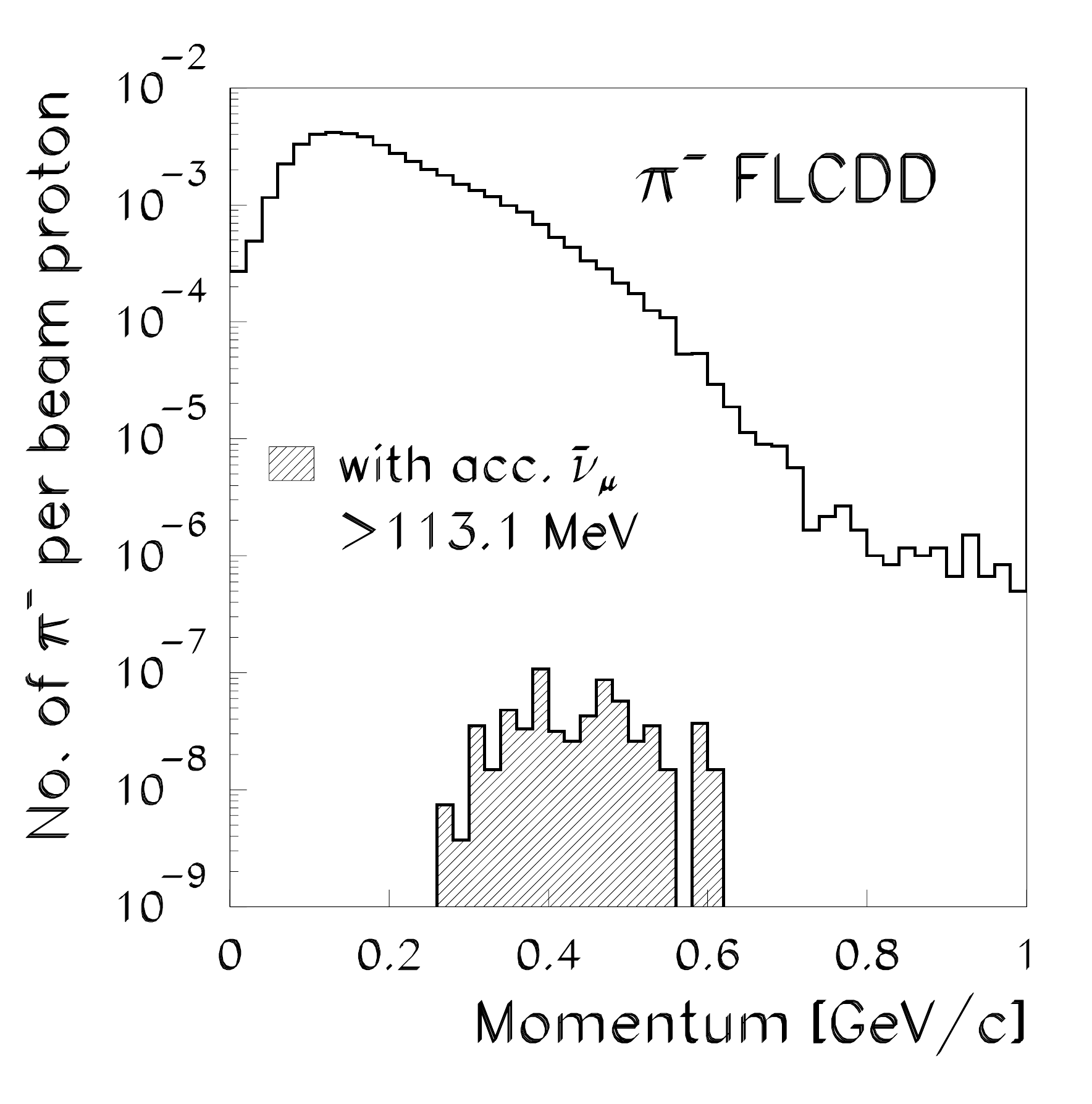} 
\caption{The `LSNDsim' prediction (FLCDD option) of 
the whole \pim\ spectrum and of the portion of the spectrum that is cut out by the
requirement of an accepted  \anum\ from  \pim\ [DIF] with energy greater than 113.1~MeV.}
\label{pimallandwithaccanumFLCDD}
\end{center}
\end{figure*}

\begin{figure*}
\begin{center}
\includegraphics[width=0.8\textwidth]{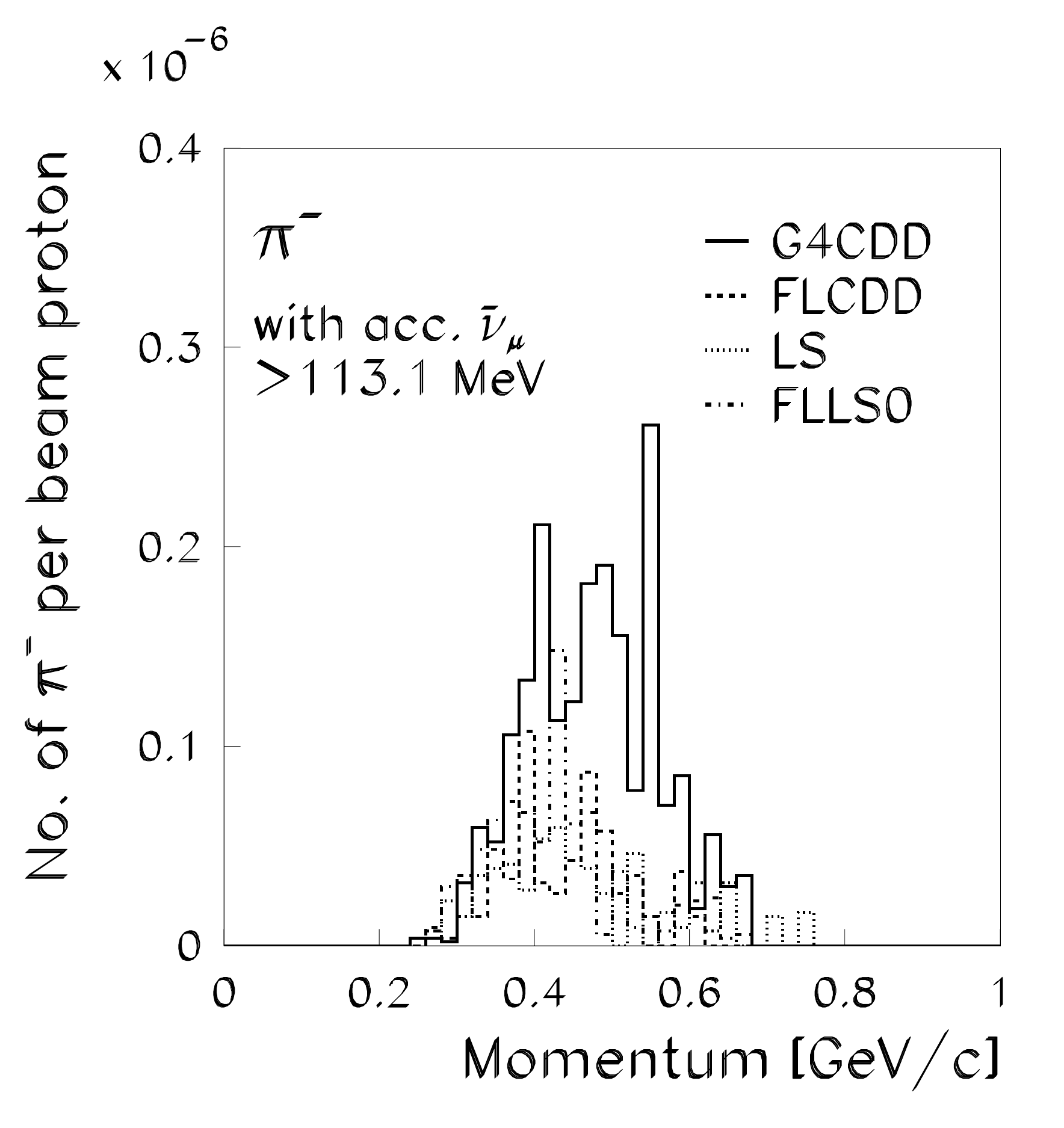} 
\caption{Predictions of the energy spectra of accepted  \anum\ from  \pim\ [DIF] with energy greater than 113.1~MeV, in the G4CDD, FLCDD, LS and FLLS0 options of `LSNDsim'.}
\label{pimacceptedanum4programs}
\end{center}
\end{figure*}

This is illustrated in Fig.~\ref{pimallandwithaccanumFLCDD} which shows for the FLCDD option of `LSNDsim'
on a logarithmic scale the whole \pim\ spectrum and the portion of the spectrum that is cut out by the
requirement of an accepted  \anum\ from  \pim\ [DIF] with energy greater than 113.1~MeV\footnote{We show the
\pim\ spectrum because the contribution of the reaction $\bar{\nu}_{\mu}$ + p $\rightarrow$ $\mu^+$ + n
to Background II is particularly large.}. The large uncertainty 
of this spectrum is highlighted in Fig.~\ref{pimacceptedanum4programs} which shows the differences between
the G4CDD, FLCDD, LS and FLLS0 options of `LSNDsim'. The uncertainty is {\em a priori\/} at 
the 100\% level\footnote{The argument that one cannot conclude from the spectrum integral on the size of the small high-energy portion of the pion spectrum can also be turned around: one cannot conclude from the high-energy portion on the size of the spectrum integral.}.
Although we apply eventually an error smaller than 100\%, for compliance with LSND's `constraints' on
neutrino fluxes that will be discussed in Section~\ref{constraints}, we consider LSND's claims unconvincing
that the numbers of \pip\ and \pim\ that decay in flight and give rise to muons
with $T_{\mu} < 3$~MeV are known with an error as small as 15\%.

LSND calculate Background II as a small fraction of all events above the respective neutrino energy thresholds. We consider that this approach renders Background II more dependent on the detailed shape of the high-energy ends of the pion spectra than necessary. We consider it safer to calculate Background II from the small portion of the neutrino spectrum between the neutrino energy threshold and 4~MeV above\footnote{The value of
4~MeV is approximate and beset with instrumental uncertainties; we bypass these 
uncertainties by considering not
absolute numbers but relative numbers: we determine a factor by dividing the average of G4CDD and FLCDD
predictions by the average of LS and FLLS0 predictions, and apply this factor as a correction to
LSND's estimate of Background II.}.

In Table~\ref{summaryofeventnumbers} we summarize the reactions and respective event numbers that
constitute Background II, and compare LSND's claims with our assessments. 
 \begin{table*}[h]
 \caption{LSND Background II event numbers}
 \label{summaryofeventnumbers}
 \begin{center}
 \begin{tabular}{|l|c|c|c|}
 \hline
 Reaction  & Background II type & \multicolumn{2}{|c|}{No. events} \\
 \hline
 \multicolumn{2}{|c|}{ }& LSND publ. & This paper \\
 \hline
 \hline
 $\bar{\nu}_{\mu}$ p $\rightarrow$$\mu^+$ n   & $T_{\mu} < 3$~MeV & $8.2$ & $10.8 \pm 8.0$ \\
 \hline
 $\nu_{\mu}$ $^{12}$C $\rightarrow$$\mu^-$ $^{12}$N  
 & $^{12}$N$^*$, $T_{\mu} < 3$~MeV, & $1.4$ & $1.8 \pm 1.8$   \\
 & $\mu^-$ capture & $-$    & $0.2 \pm 0.2$   \\
 \hline
 $\bar{\nu}_{\mu}$ $^{12}$C $\rightarrow$$\mu^+$ $^{12}$B & $T_{\mu} < 3$~MeV& $0.4$  & $0.5 \pm 0.5$   \\
 \hline
 Otherwise missed muon  & & $0.4 \pm 0.14$ & $0.4 \pm 0.14$    \\
 \hline
 $\mu^-$ $\rightarrow$ e$^-$ $\bar{\nu}_{\rm e}$ $\nu_{\mu}$, $\pi^-$ $\rightarrow$ e$^-$ $\bar{\nu}_{\rm e}$ 
   & $\bar{\nu}_{\rm e}$ events & $0.1 \pm 0.1$ & $0.1 \pm 0.1$    \\
 \hline
 \hline
Sum & & $10.5 \pm 4.6$ & $13.8 \pm  8.2$ \\
 \hline
 \end{tabular}
 \end{center}
 \end{table*}

For completeness, Table~\ref{summaryofeventnumbers} lists in addition to reactions with misidentified
low-momentum muons two more---minor and uncontroversial---contributions to Background II.

In Table~\ref{LSNDsummary} we compare Backgrounds I and II as published by LSND in Ref.~\cite{LSNDPRD64}, with the respective results obtained from `LSNDsim'. The `LSND anomaly' which is LSND's
measured `beam excess' of $117.9 \pm 22.4$ events after the subtraction of Backgrounds I and II, changes
from the published value of $87.9 \pm 23.2 $ to our value of  $73.5 \pm 25.4 $. Its significance reduces
from $3.8\:\sigma$ to $2.9\:\sigma$. 
 \begin{table*}[h]
 \caption{Backgrounds I and II  as published by LSND and as obtained from our `LSNDsim' program.}
 \label{LSNDsummary}
 \begin{center}
 \begin{tabular}{|l|c|c|}
 \hline
  & LSND published & This paper \\
 \hline
 \hline
 `Beam excess' & \multicolumn{2}{|c|}{$ 117.9 \pm 22.4$} \\
 \hline
 \hline
 Background I & $19.5 \pm 3.9 $ &  $30.6 \pm 8.8 $ \\
 \hline
 Background II & $10.5 \pm 4.6 $ &  $13.8 \pm 8.2 $ \\
 \hline
 \hline
`LSND anomaly' & $87.9 \pm 23.2 $ &  $73.5 \pm 25.4 $ \\
Significance & $3.8\:\sigma$ & $2.9\:\sigma$ \\ 
 \hline
 \end{tabular}
 \end{center}
 \end{table*}

We reiterate that in a subsequent paper~\cite{secondpaper} we shall argue that
the `LSND anomaly' and its significance is even smaller.

\section{On LSND's `constraints' on neutrino fluxes}
\label{constraints}

LSND claim that their neutrino fluxes are correct within the errors
quoted in their final physics publication~\cite{LSNDPRD64}
because certain experimental `constraints' are met. They list these `constraints'
in Table IV in Ref.~\cite{LSNDPRD64} that is reproduced here as Table~\ref{LSNDconstraints}. 
However, there is no quantitative follow-up of the `constraints'. 
\begin{table}[htdp]
\caption{``Cross section uncertainties for the neutrino reactions with two-body final states that occur in LSND. The cross sections for these processes are known accurately because either related measurements can be used to constrain the matrix elements or only fundamental particles are observed. Also shown are the corresponding neutrino flux constraints.'' (Table caption and table contents copied from Table IV in Ref.~\cite{LSNDPRD64}).}
\begin{center}
\begin{tabular}{|c|c|c|c|}
\hline
Process & $\sigma$ Constraint & $\sigma$ Uncertainty & Flux Constraint \\
\hline
\hline
$\nu$ e $\rightarrow$ $\nu$ e  & Standard model process & 1\% & 
   $\mu^+$ $\rightarrow$ $\nu_{\rm e} \bar{\nu}_{\mu} {\rm e}^+$ \: DAR \\
$^{12}$C $(\nu_{\rm e}, {\rm e}^{-})$ $^{12}$N$_{\rm gs}$ &  $^{12}$N$_{\rm gs}$ & 5\% &
   $\mu^+$ $\rightarrow$ $\nu_{\rm e} \bar{\nu}_{\mu} {\rm e}^+$ \: DAR \\ 
$^{12}$C $(\nu_{\mu}, {\mu}^{-})$ $^{12}$N$_{\rm gs}$ &  $^{12}$N$_{\rm gs}$ & 5\% &
   $\pi^+$ $\rightarrow$ $\nu_{\mu} \mu^+$ \: DIF  \\
p $(\bar{\nu}_{\mu}, \mu^{-})$ n & neutron decay & 5\% &
   $\pi^-$ $\rightarrow$ $\bar{\nu}_{\mu} \mu^-$ \: DIF  \\
\hline
\end{tabular}
\end{center}
\label{LSNDconstraints}
\end{table}

In the following, we shall argue that 
(i) our estimate of Background I from the interactions of conventional \anue , although larger by a factor of 1.6, is not in conflict with LSND's `constraints'; and that 
(ii) LSND's `constraints' concern primarily not Background I but the less important Background II  from misidentified \num\ and \anum\ interactions.

These are the cornerstones of our argumentation:
\begin{enumerate}
\item {\em On the \nue\ and \anum\ fluxes from \mup\ [DAR] decays} 

LSND claim in Ref.~\cite{LSNDNUE1997} 
that from the observed number of events from the exclusive
reaction \nue\ + $^{12}$C $\rightarrow$ \eminus\ + $^{12}$N$_{\rm gs}$, the total $\pi^+$ flux
is known to 11\% through the chain $\pi^+$ $\rightarrow$ $\mu^+$ 
$\rightarrow$ \nue . 

We are not disputing this 11\% claim. However, we note that 
LSND claim with 7\% a better precision than 11\% on the neutrino flux from \mup\ [DAR] decays,
and therefore on the total \nue\ and \anum\ fluxes. The better precision does not stem from the
said `constraint' but from the calibration experiment E866~\cite{E866} (which we dispute, see Section~\ref{critique}).

LSND claim in Table IV in Ref.~\cite{LSNDPRD64} also the elastic scattering 
process $\nu$ + e$^-$ $\rightarrow$ $\nu$ + e$^-$ to constrain the 
\nue\ and \anum\ fluxes from \mup\ [DAR] decays. We hold that the statistics are scarce and translate into a constraint on the
total $\pi^+$ flux at the 20\% level---which brings nothing new.

\item {\em On the \num\ and \anum\ fluxes from \pip\  [DIF] and \pim [DIF] decays, respectively} 

LSND claim in Ref.~\cite{LSNDNUM1997} $56.8 \pm 9.6$ (17\%) events from the exclusive
reaction \num\ + $^{12}$C $\rightarrow$ \mum\ + $^{12}$N$_{\rm gs}$ and a
measured cross-section of $(6.6 \pm 1.0 \pm 1.0) \times 10^{-41}$ cm$^2$ (21\%); for the
determination of this cross-section, the \num\ flux above the muon threshold of
123.7 MeV is assumed by LSND to be known to 15\%.

We hold that all these \num\ stem from $\pi^+$ [DIF] which is 
only 2\% of all $\pi^+$ and the increase from the claimed 7\% precision on the
number of all $\pi^+$ to the claimed 15\% precision on the number of \num\  
at the high-energy end above 123.7 MeV from \pip\ [DIF] decays
is too optimistic. We dispute LSND's claim to have measured the cross-section
of  \num\ + $^{12}$C $\rightarrow$ \mum\ + $^{12}$N$_{\rm gs}$ with 21\%
precision.
 
We turn the argument around and accept that the
measured number of events from the exclusive reaction 
\num\ + $^{12}$C $\rightarrow$ \mum\ + $^{12}$N$_{\rm gs}$ together with
its theoretically well-known cross-section (5\%)
constrains LSND's  \num\ flux above 123.7 MeV to 17\% precision.
However, we note that this precision refers to all \num\ above 123.7 MeV
but does not constrain in the same way the \num\ spectral shape as a function of energy; 
specifically, the small band between the muon threshold and 4~MeV above that is relevant
for the estimate of Background II must have
a much larger uncertainty than 17\%. Before we discuss how much larger, we consider
the `constraint' on the \anum\ flux.

LSND measured (see Table V in Ref.~\cite{LSNDNUM1997}) 
in the 1993--1995 exposure 
the sum of all events from the
reaction \anum\ + p $\rightarrow$ \mup\  + n and the
two inclusive reactions  \num\ + $^{12}$C $\rightarrow$ \mum\ X and
\anum\ + $^{12}$C $\rightarrow$ \mup\ X as $1924 \pm 47$ events,
where---according to cross-sections from theory since no experimental distinction is
possible---140 events come from
the reaction \anum\ + p $\rightarrow$ \mup\  + n, 46 events from the reaction  
\anum\ + $^{12}$C $\rightarrow$ \mup\ + X and 1738 events from the reaction
\num\ +$^{12}$C $\rightarrow$ \mum\  + X. With a 17\% precision on the
\num\ flux above the muon threshold of 123.7 MeV 
and with an estimated fraction of all events of 10.7\% due to \anum\
reactions, no constraint can be derived on the \anum\ flux.

Yet LSND claim another constraint from
the measured number of events with a muon
together with a correlated neutron from the
three reactions \anum\ + p $\rightarrow$ \mup\ + n,
\anum\ + $^{12}$C $\rightarrow$ \mup\ + n + X and
\num + $^{12}$C $\rightarrow$ \mum\ + n + X. 
This approach is motivated by the (theoretical) expectation that the requirement 
of a correlated neutron much reduces the  
preponderance of the \num\ reaction with $^{12}$C. 
The observed number of events is $210 \pm 35$ events in the 1993--1995 exposure
(see Table VI in Ref.~\cite{LSNDNUM1997}). The calculated repartition of
events among the three reactions is $210 = 140 + 36 + 34$. LSND conclude from this
that ``The observed 
number of events with neutrons
also rules out a \anum\ flux much bigger than that calculated 
by the beam Monte Carlo simulation'' 
(Section VII in Ref. ~\cite{LSNDNUM1997}).

We observe first that LSND
do not state numerically what `much bigger' means; second, the argument
depends on whether the fraction of events with a correlated neutron
is determined correctly (which we dispute); and third, the argument 
hinges on the correctness of the assumed fraction of
correlated neutrons (which is certain as 100\% for  \anum\ + p $\rightarrow$ \mup\ + n,
while predicted by theory as 60\% for \anum\ + $^{12}$C $\rightarrow$ \mup\  + X and
as 6\% for \num\ + $^{12}$C $\rightarrow$ \mum\ + X). 

With an uncertainty of the \num\ flux above 123.7 MeV of 17\%,
with an uncertainty of the \anum\ + $^{12}$C $\rightarrow$ \mup\ + X cross-section
of 100\%, and with neutron branching fractions of $(60 \pm 20)$\% and
$(6 \pm 4)$\%, respectively, there follows an uncertainty of 32\% of the \anum\
flux above 113.1 MeV. 

The \num\ and \anum\ fluxes are only relevant for the
estimation of Background II from `invisible' muons; because this background stems
only from the narrow band of neutrino energies between threshold and 4~MeV 
above threshold, and the error 
on the total fluxes above threshold
says little about the systematic errors of event numbers from neutrinos in the narrow bands, we double
and then round the errors for the fluxes in the narrow bands: 35\% for \num\ and 60\% for \anum . These
errors are used to calculate the error of Background II listed in Tables~\ref{summaryofeventnumbers}
and \ref{LSNDsummary}.

\end{enumerate}

The most important conclusion is that there is no strong constraint on the \anue\ flux from \mum\ [DAR]. 
The 32\% uncertainty of the \anum\ flux above the 113.1 MeV threshold is well consistent with
our claims of the increase by a factor of 1.6 of the conventional \anue\ rate (that stems from the integral of
all $\pi^-$ and not from the high-momentum end of $\pi^-$ decaying in flight), and of the increase of its relative error from 20\% to 29\%.

While LSND's `constraints' have little impact on the all-important conventional \anue\ rate,
they concern primarily the rate of background events with an `invisible' muon
in the final state, and its error.

\section{Critique of LSND's neutrino flux calculations}
\label{critique}

Our critique of LSND's neutrino flux calculations focusses on two aspects. The first is the inadequate 
handling of contributions from higher-generation particles in the hadronic cascade after the primary interaction of beam protons in the LSND `beam stop', primarily from neutrons. The second is the assignment of too optimistic systematic errors of the calculated neutrino fluxes (We quote: ``Calculations of \mup\ DAR fluxes are uncertain at the 7\% level, while $\pi^\pm$ 
DIF fluxes and \mum\ DAR fluxes are uncertain to 15\%'', c.f. Section II.D in Ref.~\cite{LSNDPRD64}).

The experimental pion production cross-sections that were used as input to LSND's Monte Carlo neutrino flux calculations~\cite{BurmanSmith,BurmanPlischke} were thin-target cross-sections. However, the LSND `beam
stop' represented a thick-target geometry. How was this important difference taken into account?

We quote again: ``The proton beam degradation within the target is described by the inelastic proton cross section, for which we use the energy-independent cross section measured in neutron--nucleus collisions...Each inelastic proton collision is assumed to reduce the energy of the proton by an amount appropriate for pion production...Protons are followed, losing energy either through collisional or ionization energy loss, until reaching the minimum energy for pion	production...Pion production through secondary proton interactions accounts for approximately 10\% of the total.'' (Section~2.2 in Ref.~\cite{BurmanSmith}). 

There was clearly awareness that secondary protons play a role and this was taken into account in the way
described. However, it seems that LSND underestimated the importance of hadrons up to the fourth generation, and the importance of neutrons especially for \pim\ production.

LSND claim that their Monte Carlo program was tuned to reproduce pertinent experimental data from the
E866 experiment~\cite{E866}, and use this as justification of their error assignments.

We hold that the E866 experiment was sensitive only to the integral
of \pip\ production which is rather uncontroversial. E866 claimed a 7.6\% precision on this integral.
We note that the LSND Monte Carlo program, when used for a simulation of the E866 experiment,
``was unable to reduce the predicted leakage from the side without making unacceptable changes to the
pion decay inside the stack'' (Section 3 in Ref.~\cite{BurmanSmith}). The discrepancy between observed
and predicted pion leakage, apparent in Fig.~11 of Ref.~\cite{BurmanSmith} suggests that even after adjustment
of overall normalization, the LSND Monte Carlo exhibits serious shortcomings in the predictions of the momentum dependence of pion spectra. As a consequence, the systematic precision of the \num\ flux from stopped \pip\ must be worse than the 7.6\% precision claimed by E866.

We hold further that the E866 experiment is hardly sensitive to the \pim\ component of the hadronic cascade,
so there is from E866 no experimental check of LSND's claims on the amount and the uncertainty of 
\pim\ production in the LSND `beam stop'.

\section{Conclusion}

For the importance of the `LSND anomaly' that is in stark conflict with the Standard Model, we have undertaken 
a re-evaluation of the data analysis of the LSND experiment. We compare an emulation of this experiment
where we try to reproduce closely the LSND data analysis, with results of a simulation where
the model of hadron production is considerably more realistic. We use new experimental cross-sections
of pion production by protons from HARP--CDP, and four decades old data on pion production by neutrons.
The more realistic model of hadron production leads to a considerably larger conventional \anue\ flux
than was estimated by LSND. Taking merely into account the more realistic model of hadron production, the significance of the `LSND anomaly' reduces from 3.8~$\sigma$ to 2.9~$\sigma$.

\section*{Acknowledgements}

We are greatly indebted to many technical collaborators whose 
diligent and hard work made the HARP detector a well-functioning 
instrument. We thank all HARP colleagues who devoted time and 
effort to the design and construction of the detector, to data taking, 
and to setting up the computing and software infrastructure. 
We express our sincere gratitude to HARP's funding agencies 
for their support.

\end{document}

%% file: Table800MeVproH2O.tex
 \begin{table*}[h]
 \begin{scriptsize}
 \caption{Double-differential inclusive cross-section ${\rm d}^2 \sigma /{\rm d}p{\rm d}\Omega$ [mb/(GeV/{\it c} sr)] of the production of pions
  in p + H$_2$O $\rightarrow$ $\pi^\pm$ + X interactions
  with $1.5$~GeV/{\it c} beam momentum; the first error is statistical, the second systematic; $p_{\rm T}$ in GeV/{\it c}, polar angle $\theta$ in degrees.}
 \label{pro.H2O}
 \begin{center}
 \begin{tabular}{|c||c|c|}
 \hline
  & \multicolumn{2}{c|}{$20<\theta<60$} \\
 \hline
 $p_{\rm T}$ & {$\pi^+$} & {$\pi^-$} \\
 \hline
 \hline
$0.10-0.15$ &  33.6 $\pm$    1.9  $\pm$    2.1 &   7.0  $\pm$    0.9  $\pm$    0.4 \\
$0.15-0.20$ &  53.7 $\pm$    2.3  $\pm$    3.3 &   6.7  $\pm$    0.8  $\pm$    0.4 \\
$0.20-0.25$ &  47.2 $\pm$    2.1  $\pm$    2.9 &   3.5  $\pm$    0.6  $\pm$    0.2 \\
$0.25-0.35$ &  15.4 $\pm$    0.9  $\pm$    0.9 &   1.1  $\pm$    0.2  $\pm$    0.1 \\
 \hline
 \hline
  & \multicolumn{2}{c|}{$60<\theta<125$} \\
 \hline
 $p_{\rm T}$  & {$\pi^+$} & {$\pi^-$} \\
 \hline
 \hline
$0.10-0.15$ &  23.9 $\pm$    1.3  $\pm$    1.5 &   5.6  $\pm$    0.7  $\pm$    0.3 \\
$0.15-0.20$ &  32.4 $\pm$    1.4  $\pm$    2.0 &   5.4  $\pm$    0.6  $\pm$    0.3 \\
$0.20-0.25$ &  21.5 $\pm$    1.2  $\pm$    1.3 &   3.2  $\pm$    0.4  $\pm$    0.2 \\
$0.25-0.35$ &   7.5 $\pm$    0.5  $\pm$    0.5 &   0.9  $\pm$    0.2  $\pm$    0.1 \\
 \hline
 \end{tabular}
 \end{center}
 \end{scriptsize}
 \end{table*}

%% file: Table800MeVpipH2O.tex
 \begin{table*}[h]
 \begin{scriptsize}
 \caption{Double-differential inclusive cross-section ${\rm d}^2 \sigma /{\rm d}p{\rm d}\Omega$ [mb/(GeV/{\it c} sr)] of the production of pions
  in $\pi^+$ + H$_2$O $\rightarrow$ $\pi^\pm$ + X interactions
  with $1.5$~GeV/{\it c} beam momentum; the first error is statistical, the second systematic; $p_{\rm T}$ in GeV/{\it c}, polar angle $\theta$ in degrees.}
 \label{pip.H2O}
 \begin{center}
 \begin{tabular}{|c||c|c|}
 \hline
  & \multicolumn{2}{c|}{$20<\theta<60$} \\
 \hline
 $p_{\rm T}$ & {$\pi^+$} & {$\pi^-$} \\
 \hline
 \hline
$0.10-0.15$ &  51.8 $\pm$    1.7  $\pm$    3.3 &  29.2  $\pm$    1.3  $\pm$    1.9 \\
$0.15-0.20$ &  70.1 $\pm$    1.8  $\pm$    4.5 &  26.8  $\pm$    1.1  $\pm$    1.7 \\
$0.20-0.25$ &  71.8 $\pm$    1.8  $\pm$    4.6 &  26.7  $\pm$    1.1  $\pm$    1.7 \\
$0.25-0.35$ &  59.4 $\pm$    1.2  $\pm$    3.8 &  22.5  $\pm$    0.7  $\pm$    1.5 \\
$0.35-0.45$ &  50.9 $\pm$    1.2  $\pm$    3.3 &  13.8  $\pm$    0.6  $\pm$    0.9 \\
 \hline
 \hline
  & \multicolumn{2}{c|}{$60<\theta<125$} \\
 \hline
 $p_{\rm T}$  & {$\pi^+$} & {$\pi^-$} \\
 \hline
 \hline
$0.10-0.15$ &  43.4 $\pm$    1.3  $\pm$    2.8 &  23.1  $\pm$    1.0  $\pm$    1.5 \\
$0.15-0.20$ &  50.0 $\pm$    1.3  $\pm$    3.2 &  22.1  $\pm$    0.8  $\pm$    1.4 \\
$0.20-0.25$ &  39.0 $\pm$    1.1  $\pm$    2.5 &  16.3  $\pm$    0.7  $\pm$    1.1 \\
$0.25-0.35$ &  24.1 $\pm$    0.6  $\pm$    1.6 &  10.0  $\pm$    0.4  $\pm$    0.6 \\
$0.35-0.45$ &  18.9 $\pm$    0.6  $\pm$    1.2 &   5.1  $\pm$    0.3  $\pm$    0.3 \\
 \hline
 \end{tabular}
 \end{center}
 \end{scriptsize}
 \end{table*}

%% file: Table800MeVproCu.tex
 \begin{table*}[h]
 \begin{scriptsize}
 \caption{Double-differential inclusive cross-section ${\rm d}^2 \sigma /{\rm d}p{\rm d}\Omega$ [mb/(GeV/{\it c} sr)] of the production of pions
  in p + Cu $\rightarrow$ $\pi^\pm$ + X interactions
  with $1.5$~GeV/{\it c} beam momentum; the first error is statistical, the second systematic; $p_{\rm T}$ in GeV/{\it c}, polar angle $\theta$ in degrees.}
 \label{pro.Cu}
 \begin{center}
 \begin{tabular}{|c||c|c|}
 \hline
  & \multicolumn{2}{c|}{$20<\theta<60$} \\
 \hline
 $p_{\rm T}$ & {$\pi^+$} & {$\pi^-$} \\
 \hline
 \hline
$0.10-0.15$ &  43.8 $\pm$    5.9  $\pm$    3.2 &  22.8  $\pm$    4.4  $\pm$    1.7 \\
$0.15-0.20$ &  53.2 $\pm$    5.8  $\pm$    3.9 &  20.2  $\pm$    3.6  $\pm$    1.5 \\
$0.20-0.25$ &  42.1 $\pm$    5.2  $\pm$    3.1 &   8.2  $\pm$    2.3  $\pm$    0.6 \\
$0.25-0.35$ &  19.8 $\pm$    2.5  $\pm$    1.4 &   1.7  $\pm$    0.7  $\pm$    0.1 \\
 \hline
 \hline
  & \multicolumn{2}{c|}{$60<\theta<125$} \\
 \hline
 $p_{\rm T}$  & {$\pi^+$} & {$\pi^-$} \\
 \hline
 \hline
$0.15-0.20$ &  36.2 $\pm$    4.0  $\pm$    4.0 &  10.5  $\pm$    2.1  $\pm$    1.2 \\
$0.20-0.25$ &  22.5 $\pm$    2.9  $\pm$    2.5 &   4.7  $\pm$    1.3  $\pm$    0.5 \\
$0.25-0.35$ &   8.6 $\pm$    1.2  $\pm$    1.0 &   2.3  $\pm$    0.6  $\pm$    0.3 \\
 \hline
 \end{tabular}
 \end{center}
 \end{scriptsize}
 \end{table*}

%% file: Table800MeVpipCu.tex
 \begin{table*}[h]
 \begin{scriptsize}
 \caption{Double-differential inclusive cross-section ${\rm d}^2 \sigma /{\rm d}p{\rm d}\Omega$ [mb/(GeV/{\it c} sr)] of the production of pions
  in $\pi^+$ + Cu $\rightarrow$ $\pi^\pm$ + X interactions
  with $1.5$~GeV/{\it c} beam momentum; the first error is statistical, the second systematic; $p_{\rm T}$ in GeV/{\it c}, polar angle $\theta$ in degrees.}
 \label{pip.Cu}
 \begin{center}
 \begin{tabular}{|c||c|c|}
 \hline
  & \multicolumn{2}{c|}{$20<\theta<60$} \\
 \hline
 $p_{\rm T}$ & {$\pi^+$} & {$\pi^-$} \\
 \hline
 \hline
$0.10-0.15$ &  96.2 $\pm$    6.0  $\pm$    7.3 &  58.9  $\pm$    4.6  $\pm$    4.4 \\
$0.15-0.20$ & 110.0 $\pm$    5.9  $\pm$    8.3 &  54.0  $\pm$    4.0  $\pm$    4.1 \\
$0.20-0.25$ &  96.3 $\pm$    5.3  $\pm$    7.3 &  49.0  $\pm$    3.8  $\pm$    3.7 \\
$0.25-0.35$ &  86.7 $\pm$    3.5  $\pm$    6.5 &  37.1  $\pm$    2.3  $\pm$    2.8 \\
$0.35-0.45$ &  78.7 $\pm$    3.6  $\pm$    5.9 &  27.7  $\pm$    2.1  $\pm$    2.1 \\
 \hline
 \hline
  & \multicolumn{2}{c|}{$60<\theta<125$} \\
 \hline
 $p_{\rm T}$  & {$\pi^+$} & {$\pi^-$} \\
 \hline
 \hline
$0.15-0.20$ &  92.5 $\pm$    4.3  $\pm$   10.5 &  54.2  $\pm$    3.3  $\pm$    6.2 \\
$0.20-0.25$ &  74.5 $\pm$    3.7  $\pm$    8.5 &  40.0  $\pm$    2.7  $\pm$    4.5 \\
$0.25-0.35$ &  41.8 $\pm$    1.9  $\pm$    4.7 &  21.8  $\pm$    1.4  $\pm$    2.5 \\
$0.35-0.45$ &  29.8 $\pm$    1.7  $\pm$    3.4 &  13.1  $\pm$    1.1  $\pm$    1.5 \\
 \hline
 \end{tabular}
 \end{center}
 \end{scriptsize}
 \end{table*}

%% file: Table800MeVproTa.tex
 \begin{table*}[h]
 \begin{scriptsize}
 \caption{Double-differential inclusive cross-section ${\rm d}^2 \sigma /{\rm d}p{\rm d}\Omega$ [mb/(GeV/{\it c} sr)] of the production of pions
  in p + Ta $\rightarrow$ $\pi^\pm$ + X interactions
  with $1.5$~GeV/{\it c} beam momentum; the first error is statistical, the second systematic; $p_{\rm T}$ in GeV/{\it c}, polar angle $\theta$ in degrees.}
 \label{pro.Ta}
 \begin{center}
 \begin{tabular}{|c||c|c|}
 \hline
  & \multicolumn{2}{c|}{$20<\theta<60$} \\
 \hline
 $p_{\rm T}$ & {$\pi^+$} & {$\pi^-$} \\
 \hline
 \hline
$0.10-0.15$ &  60.6 $\pm$   10.4  $\pm$    4.4 &  26.3  $\pm$    6.6  $\pm$    1.9 \\
$0.15-0.20$ &  70.6 $\pm$   10.5  $\pm$    5.1 &  33.9  $\pm$    7.3  $\pm$    2.5 \\
$0.20-0.25$ &  63.1 $\pm$    9.8  $\pm$    4.6 &  16.7  $\pm$    5.1  $\pm$    1.2 \\
$0.25-0.35$ &  24.9 $\pm$    4.3  $\pm$    1.8 &   5.2  $\pm$    2.0  $\pm$    0.4 \\
 \hline
 \hline
  & \multicolumn{2}{c|}{$60<\theta<125$} \\
 \hline
 $p_{\rm T}$  & {$\pi^+$} & {$\pi^-$} \\
 \hline
 \hline
$0.15-0.20$ &  46.0 $\pm$    7.4  $\pm$    5.1 &  23.3  $\pm$    5.4  $\pm$    2.6 \\
$0.20-0.25$ &  34.6 $\pm$    5.9  $\pm$    3.9 &  11.3  $\pm$    3.4  $\pm$    1.3 \\
$0.25-0.35$ &  12.2 $\pm$    2.3  $\pm$    1.4 &   3.3  $\pm$    1.2  $\pm$    0.4 \\
 \hline
 \end{tabular}
 \end{center}
 \end{scriptsize}
 \end{table*}

%% file: Table800MeVpipTa.tex
 \begin{table*}[h]
 \begin{scriptsize}
 \caption{Double-differential inclusive cross-section ${\rm d}^2 \sigma /{\rm d}p{\rm d}\Omega$ [mb/(GeV/{\it c} sr)] of the production of pions
  in $\pi^+$ + Ta $\rightarrow$ $\pi^\pm$ + X interactions
  with $1.5$~GeV/{\it c} beam momentum; the first error is statistical, the second systematic; $p_{\rm T}$ in GeV/{\it c}, polar angle $\theta$ in degrees.}
 \label{pip.Ta}
 \begin{center}
 \begin{tabular}{|c||c|c|}
 \hline
  & \multicolumn{2}{c|}{$20<\theta<60$} \\
 \hline
 $p_{\rm T}$ & {$\pi^+$} & {$\pi^-$} \\
 \hline
 \hline
$0.10-0.15$ & 138.3 $\pm$   11.9  $\pm$   10.4 &  98.1  $\pm$    9.5  $\pm$    7.4 \\
$0.15-0.20$ & 137.2 $\pm$   10.2  $\pm$   10.4 &  93.1  $\pm$    8.3  $\pm$    7.0 \\
$0.20-0.25$ & 132.3 $\pm$    9.6  $\pm$   10.0 &  68.3  $\pm$    6.9  $\pm$    5.2 \\
$0.25-0.35$ & 114.8 $\pm$    6.4  $\pm$    8.7 &  65.8  $\pm$    4.9  $\pm$    5.0 \\
$0.35-0.45$ & 117.7 $\pm$    6.8  $\pm$    8.9 &  48.3  $\pm$    4.4  $\pm$    3.6 \\
 \hline
 \hline
  & \multicolumn{2}{c|}{$60<\theta<125$} \\
 \hline
 $p_{\rm T}$  & {$\pi^+$} & {$\pi^-$} \\
 \hline
 \hline
$0.15-0.20$ & 142.0 $\pm$    8.9  $\pm$   16.1 &  93.8  $\pm$    7.6  $\pm$   10.7 \\
$0.20-0.25$ & 123.4 $\pm$    7.8  $\pm$   14.0 &  56.4  $\pm$    5.1  $\pm$    6.4 \\
$0.25-0.35$ &  67.6 $\pm$    4.0  $\pm$    7.7 &  35.4  $\pm$    2.9  $\pm$    4.0 \\
$0.35-0.45$ &  45.9 $\pm$    3.4  $\pm$    5.2 &  22.5  $\pm$    2.3  $\pm$    2.6 \\
 \hline
 \end{tabular}
 \end{center}
 \end{scriptsize}
 \end{table*}

%% file: Table800MeVproPb.tex
 \begin{table*}[h]
 \begin{scriptsize}
 \caption{Double-differential inclusive cross-section ${\rm d}^2 \sigma /{\rm d}p{\rm d}\Omega$ [mb/(GeV/{\it c} sr)] of the production of pions
  in p + Pb $\rightarrow$ $\pi^\pm$ + X interactions
  with $1.5$~GeV/{\it c} beam momentum; the first error is statistical, the second systematic; $p_{\rm T}$ in GeV/{\it c}, polar angle $\theta$ in degrees.}
 \label{pro.Pb}
 \begin{center}
 \begin{tabular}{|c||c|c|}
 \hline
  & \multicolumn{2}{c|}{$20<\theta<60$} \\
 \hline
 $p_{\rm T}$ & {$\pi^+$} & {$\pi^-$} \\
 \hline
 \hline
$0.10-0.15$ &  49.1 $\pm$    9.8  $\pm$    3.6 &  44.5  $\pm$    9.8  $\pm$    3.2 \\
$0.15-0.20$ &  54.2 $\pm$    9.6  $\pm$    3.9 &  35.3  $\pm$    7.9  $\pm$    2.6 \\
$0.20-0.25$ &  56.0 $\pm$   10.2  $\pm$    4.1 &  12.9  $\pm$    4.9  $\pm$    0.9 \\
$0.25-0.35$ &  23.2 $\pm$    4.6  $\pm$    1.7 &   6.6  $\pm$    2.3  $\pm$    0.5 \\
 \hline
 \hline
  & \multicolumn{2}{c|}{$60<\theta<125$} \\
 \hline
 $p_{\rm T}$  & {$\pi^+$} & {$\pi^-$} \\
 \hline
 \hline
$0.15-0.20$ &  41.4 $\pm$    7.2  $\pm$    4.6 &  30.5  $\pm$    6.1  $\pm$    3.4 \\
$0.20-0.25$ &  43.4 $\pm$    6.8  $\pm$    4.9 &   9.0  $\pm$    3.0  $\pm$    1.0 \\
$0.25-0.35$ &  13.5 $\pm$    2.4  $\pm$    1.5 &   4.2  $\pm$    1.4  $\pm$    0.5 \\
 \hline
 \end{tabular}
 \end{center}
 \end{scriptsize}
 \end{table*}

%% file: Table800MeVpipPb.tex
 \begin{table*}[h]
 \begin{scriptsize}
 \caption{Double-differential inclusive cross-section ${\rm d}^2 \sigma /{\rm d}p{\rm d}\Omega$ [mb/(GeV/{\it c} sr)] of the production of pions
  in $\pi^+$ + Pb $\rightarrow$ $\pi^\pm$ + X interactions
  with $1.5$~GeV/{\it c} beam momentum; the first error is statistical, the second systematic; $p_{\rm T}$ in GeV/{\it c}, polar angle $\theta$ in degrees.}
 \label{pip.Pb}
 \begin{center}
 \begin{tabular}{|c||c|c|}
 \hline
  & \multicolumn{2}{c|}{$20<\theta<60$} \\
 \hline
 $p_{\rm T}$ & {$\pi^+$} & {$\pi^-$} \\
 \hline
 \hline
$0.10-0.15$ & 134.8 $\pm$   11.9  $\pm$   10.2 & 100.4  $\pm$   10.1  $\pm$    7.6 \\
$0.15-0.20$ & 136.2 $\pm$   10.2  $\pm$   10.3 &  81.9  $\pm$    8.1  $\pm$    6.2 \\
$0.20-0.25$ & 152.4 $\pm$   11.0  $\pm$   11.5 &  72.9  $\pm$    7.5  $\pm$    5.5 \\
$0.25-0.35$ & 125.4 $\pm$    6.9  $\pm$    9.5 &  68.7  $\pm$    5.1  $\pm$    5.2 \\
$0.35-0.45$ & 131.2 $\pm$    7.7  $\pm$    9.9 &  58.2  $\pm$    5.0  $\pm$    4.4 \\
 \hline
 \hline
  & \multicolumn{2}{c|}{$60<\theta<125$} \\
 \hline
 $p_{\rm T}$  & {$\pi^+$} & {$\pi^-$} \\
 \hline
 \hline
$0.15-0.20$ & 149.0 $\pm$    9.5  $\pm$   16.9 &  96.9  $\pm$    7.5  $\pm$   11.0 \\
$0.20-0.25$ & 104.9 $\pm$    7.3  $\pm$   11.9 &  81.6  $\pm$    6.5  $\pm$    9.3 \\
$0.25-0.35$ &  69.6 $\pm$    4.1  $\pm$    7.9 &  40.0  $\pm$    3.2  $\pm$    4.5 \\
$0.35-0.45$ &  53.9 $\pm$    3.7  $\pm$    6.1 &  25.3  $\pm$    2.5  $\pm$    2.9 \\
 \hline
 \end{tabular}
 \end{center}
 \end{scriptsize}
 \end{table*}

%% file: Table800MeVratio.tex
 \begin{table*}[h]
 \begin{scriptsize}
 \caption{Ratio of inclusive double-differential cross-sections ${\rm d}^2 \sigma /{\rm d}p{\rm d}\Omega$ of $\pi^-$ to $\pi^+$ production by $1.5$~GeV/{\it c} protons in a 60~cm thick water target; the first error is statistical, the second systematic; $p_{\rm T}$ in GeV/{\it c}, polar angle $\theta$ in degrees, fiducial $z$ range in target in cm.}
 \label{ratiointhickwater}
 \begin{center}
 \begin{tabular}{|c||c|c|}
 \hline
 $p_{\rm T}$ & {$20<\theta<35$} and {$3<z<56$} & {$35<\theta<50$} and {$3<z<59$} \\
 \hline
 \hline
$0.100-0.125$ & 0.203  $\pm$  0.018  $\pm$  0.008 & 0.270  $\pm$  0.028  $\pm$  0.023 \\
$0.125-0.150$ & 0.112  $\pm$  0.010  $\pm$  0.007 & 0.210  $\pm$  0.018  $\pm$  0.008 \\
$0.150-0.175$ & 0.119  $\pm$  0.010  $\pm$  0.005 & 0.169  $\pm$  0.014  $\pm$  0.010 \\
$0.175-0.200$ & 0.065  $\pm$  0.007  $\pm$  0.004 & 0.126  $\pm$  0.011  $\pm$  0.005 \\
$0.200-0.225$ & 0.041  $\pm$  0.006  $\pm$  0.002 & 0.101  $\pm$  0.010  $\pm$  0.007 \\
$0.225-0.250$ & 0.064  $\pm$  0.011  $\pm$  0.003 & 0.072  $\pm$  0.008  $\pm$  0.003 \\
$0.250-0.275$ & 0.074  $\pm$  0.015  $\pm$  0.010 & 0.080  $\pm$  0.011  $\pm$  0.003 \\
$0.275-0.300$ & 0.076  $\pm$  0.021  $\pm$  0.013 & 0.077  $\pm$  0.013  $\pm$  0.006 \\
 \hline
 $p_{\rm T}$ & {$50<\theta<65$} and {$3<z<59$} & {$65<\theta<80$} and {$6<z<59$} \\
 \hline
 \hline
$0.100-0.125$ & 0.235  $\pm$  0.028  $\pm$  0.008 & 0.227  $\pm$  0.029  $\pm$  0.023 \\
$0.125-0.150$ & 0.227  $\pm$  0.020  $\pm$  0.017 & 0.214  $\pm$  0.021  $\pm$  0.007 \\
$0.150-0.175$ & 0.198  $\pm$  0.017  $\pm$  0.008 & 0.174  $\pm$  0.016  $\pm$  0.006 \\
$0.175-0.200$ & 0.145  $\pm$  0.013  $\pm$  0.008 & 0.157  $\pm$  0.015  $\pm$  0.010 \\
$0.200-0.225$ & 0.111  $\pm$  0.011  $\pm$  0.004 & 0.112  $\pm$  0.013  $\pm$  0.006 \\
$0.225-0.250$ & 0.109  $\pm$  0.012  $\pm$  0.004 & 0.094  $\pm$  0.012  $\pm$  0.006 \\
$0.250-0.275$ & 0.098  $\pm$  0.013  $\pm$  0.004 & 0.091  $\pm$  0.015  $\pm$  0.003 \\
$0.275-0.300$ & 0.100  $\pm$  0.016  $\pm$  0.015 & 0.137  $\pm$  0.023  $\pm$  0.004 \\
 \hline
 $p_{\rm T}$ & {$80<\theta<95$} and {$11<z<59$} & {$95<\theta<125$} and {$31<z<59$} \\
 \hline
 \hline
$0.100-0.125$ & 0.326  $\pm$  0.043  $\pm$  0.021 & 0.240  $\pm$  0.027  $\pm$  0.009 \\
$0.125-0.150$ & 0.214  $\pm$  0.022  $\pm$  0.011 & 0.201  $\pm$  0.020  $\pm$  0.011 \\
$0.150-0.175$ & 0.168  $\pm$  0.017  $\pm$  0.008 & 0.141  $\pm$  0.018  $\pm$  0.009 \\
$0.175-0.200$ & 0.127  $\pm$  0.015  $\pm$  0.007 & 0.129  $\pm$  0.018  $\pm$  0.006 \\
$0.200-0.225$ & 0.112  $\pm$  0.014  $\pm$  0.004 & 0.168  $\pm$  0.030  $\pm$  0.006 \\
$0.225-0.250$ & 0.114  $\pm$  0.018  $\pm$  0.004 & 0.099  $\pm$  0.027  $\pm$  0.023 \\
$0.250-0.275$ & 0.114  $\pm$  0.022  $\pm$  0.004 & 0.151  $\pm$  0.047  $\pm$  0.009 \\
$0.275-0.300$ & 0.153  $\pm$  0.035  $\pm$  0.008 & 0.227  $\pm$  0.079  $\pm$  0.021 \\
 \hline
 \end{tabular}
 \end{center}
 \end{scriptsize}
 \end{table*}